\documentclass[journal]{IEEEtran}
\usepackage{graphicx}
\usepackage{subcaption}
\usepackage{multirow}
\usepackage{amsmath}
\usepackage{amsfonts}
\usepackage{booktabs}
\usepackage{graphicx}
\usepackage{multicol}
\usepackage{multirow}
\usepackage{rotating}
\DeclareGraphicsExtensions{.pdf}

\input{Qcircuit.tex}

\begin{document}
%\onecolumn
%
% paper title
\title{ T-count Optimized Design of Quantum Integer Multiplication}
%
%
% author names and IEEE memberships
% note positions of commas and nonbreaking spaces ( ~ ) LaTeX will not break
% a structure at a ~ so this keeps an author's name from being broken across
% two lines.
% use \thanks{} to gain access to the first footnote area
% a separate \thanks must be used for each paragraph as LaTeX2e's \thanks
% was not built to handle multiple paragraphs
\author{Edgard Mu\~{n}oz-Coreas, Himanshu Thapliyal%~\IEEEmembership{ Member,~IEEE}
\thanks{  }%
\thanks{Edgard Mu\~{n}oz-Coreas and Himanshu Thapliyal are with Department of Electrical and Computer Engineering, University of Kentucky, Lexington, KY, USA. \newline
Email :hthapliyal@uky.edu }}
\maketitle

%\begin{multicols}{2}

\begin{abstract}
Quantum circuits of many qubits are extremely difficult to realize; thus, the number of qubits is an important metric in a quantum circuit design.  Further, scalable and reliable  quantum circuits are based on Clifford + T gates.  An efficient quantum circuit saves quantum hardware resources by reducing the number of T gates without substantially increasing the number of qubits.  Recently, the design of a quantum multiplier is presented by Babu \cite{Babu} which improves the existing works in terms of number of quantum gates, number of qubits, and delay.  However, the recent design is not based on fault-tolerant Clifford + T gates. Also, it has large number of qubits and garbage outputs.  Therefore, this work presents a T-count optimized quantum circuit for integer multiplication with only $4 \cdot n + 1$ qubits and no garbage outputs.  The proposed quantum multiplier design saves the T-count by using a novel quantum conditional adder circuit. Also, where one operand to the controlled adder is zero,  the conditional adder is replaced with a Toffoli gate array to further save the T gates.  To have fair comparison with the recent design by Babu and get an actual  estimate of the T-count, it is made garbageless by using Bennett's garbage removal scheme.   The proposed design achieves an average T-count savings of  $47.55 \%$ compared to the recent work by Babu.  Further,  comparison is also performed with other recent works  by Lin et. al. \cite{Lin}, and Jayashree et. al.\cite{Jayashree}.  Average T-count savings of  $62.71 \%$ and $26.30$ \% are achieved compared to the recent works by Lin et. al., and Jayashree et. al., respectively.              

\end{abstract}

\section{Introduction}
\label{md_start}

Among the emerging computing paradigms, quantum computing appears to be promising due to its applications in number 
theory, encryption, search and scientific computation \cite{Cheung} \cite{Bowregard} \cite{Montanaro} \cite{Shparlinski} \cite{Proos} \cite{Seroussi}.  Quantum circuits for integer arithmetic operations such as addition, subtraction and multiplication are required in the quantum circuit implementations of many quantum algorithms in these areas.  Thus, researchers have included dedicated libraries of basic quantum integer arithmetic functions for use in quantum programming languages such as Quipper and LIQUi and in quantum computing design tools such as those proposed in \cite{Quipper} \cite{LIQUi} \cite{Maslov34} and \cite{Haner}.

  Quantum circuits do not lose information during computation and quantum computation can only be performed when the system consists of quantum gates.  Thus, in any Quantum circuit there is a one-to-one mapping between the input and output vectors.  Any constant inputs in the quantum circuit are called ancillae.  Garbage output refers to any output which exists in the quantum circuit to preserve one-to-one mapping but is not one of the primary inputs nor a useful output.  The inputs regenerated at the outputs are not considered garbage outputs \cite{Fredkin}.  Ancillae and garbage outputs are circuit overhead that need to be minimized.

The fault tolerant implementation of quantum circuits is gaining the attention of researchers because physical quantum computers are prone to noise errors \cite{Webster} \cite{Zhou} \cite{Paler_DAC} \cite{Polian_DAC}.  Fault tolerant implementations of quantum gates and quantum error correcting codes can be used to overcome the limits imposed by noise errors in implementing quantum computing \cite{Zhou} \cite{Mosca2}.  Recently, researchers have implemented quantum logic gates such as the controlled phase gate, controlled square-root-of-not gate, Toffoli gate, Fredkin gate and quantum full adder with the fault tolerant Clifford + T gate set due to its demonstrated tolerance to noise errors \cite{Miller} \cite{Maslov}.  However, the increased tolerance to noise errors comes with the increased implementation overhead associated with the quantum T gate \cite{Mosca2} \cite{Maslov}.  Because of the increased cost to realize the T gate, T-count has become an important performance measure for fault tolerant quantum circuit design \cite{Mosca2} \cite{Gosset}.

The design of quantum integer multiplication circuits has received notable attention in the literature.  Garbage-less designs such as those in \cite{Lidia}, \cite{Lin} have significant T gate costs.  Other works such 
as the recent design in \cite{Babu} present T gate efficient designs but do not include the additional ancillae and T gate 
costs from eliminating garbage outputs in the cost calculations.  As a result, the total quantum circuit may end up requiring 
$n$ extra qubits and the actual T gate cost may end up being doubled.  While the integer multiplication circuits presented in 
existing works such as \cite{Lin} and \cite{Babu} are interesting designs, these integer multipliers have a significant gate 
overhead in terms of T-count.  \textit{In this work, we present the design of a quantum integer multiplication circuit that is 
garbageless, requires $4 \cdot n + 1$ qubits and is optimized for T-count.  The quantum integer multiplication circuit is 
based on a proposed quantum conditional addition circuit with no input carry that that is garbageless and optimized for T-
count.}  The proposed quantum integer multiplication circuit based on our proposed design is compared and is shown to be 
better than existing designs of quantum integer multiplication circuit in terms of T-count.

This paper is organized as follows. Section \ref{md_ref} presents background information on the Clifford + T fault tolerant quantum gate family, defines the T-count performance measure, presents the algorithm that that the proposed quantum integer multiplication circuit is based on and describes the Bennett's garbage removal scheme.  In section \ref{md_adder} the design of the proposed quantum conditional addition circuit with no input carry is discussed.  The design of the proposed quantum integer multiplication circuit is presented in section \ref{md_mult}.

\section{Background}
\label{md_ref}

%\begin{figure}[h]
%\small
%\flushleft
%	\begin{subfigure}[hb]{1.5in}
%	\flushleft
%		\[
%		\Qcircuit @C=1em @R=.7em {
%		\lstick{A} & \ctrl{1} & \qw & \rstick{A}\\
%		\lstick{B} & \targ & \qw & \rstick{A \oplus B} \\
%		} 
%		\]
%	\caption{Feynman Gate}
%	\end{subfigure}  \qquad \begin{subfigure}[hb]{1.5in}
%	\flushleft
%		\[
%		\Qcircuit @C = 1em @R = .7em @!R{
%		\lstick{A} & \ctrl{1} & \qw & \rstick{A}\\
%		\lstick{B} & \ctrl{1} & \qw & \rstick{B}\\
%		\lstick{C} & \targ & \qw & \rstick{A \cdot B \oplus C} \\
%		}
%		\]
%	\caption{Toffoli Gate}	
%	\end{subfigure} 
%		
%\caption{    }
%\label{md-fig:1}
%\end{figure}

\begin{table}[tbhp]
\centering

%\begin{tabular}{ |p{5cm}||p{4cm}|p{3cm}| }
\begin{tabular}{|c|c|c|}
\hline

\textbf{ Type of Gate} &\textbf{Symbol} & \textbf{Matrix}\\
 \hline
 Not gate  & $N$   &    $\begin{bmatrix}
    0 & 1  \\
    1 & 0 
  \end{bmatrix}$ \\
 
 \hline
 Hadamard gate   & $H$   &   $ \frac{1}{\sqrt{2}}
  \begin{bmatrix}
    1 & 1  \\
    1 & -1 
  \end{bmatrix} $\\
  
\hline
$T$ gate  & $T$   &  $\begin{bmatrix}
    1 & 0  \\
    0 & e^{i.\frac{\pi}{4}} 
  \end{bmatrix} $\\ 
 
\hline
$T$ gate Hermitian transpose  & $T^{\dag}$   &  $\begin{bmatrix}
    1 & 0  \\
    0 & e^{-i.\frac{\pi}{4}} 
  \end{bmatrix} $\\  
 
 \hline
 Phase gate & $S$   &  $  \begin{bmatrix}
    1 & 0  \\
    0 & i 
  \end{bmatrix}$\\
 \hline

 Phase gate Hermitian transpose  & $S^{\dag}$   &  $  \begin{bmatrix}
    1 & 0  \\
    0 & -i 
  \end{bmatrix}$\\
 \hline
 
Feynman gate  & $C$    & $\begin{bmatrix}
    1 & 0 & 0 & 0 \\
    0 & 1 & 0 & 0\\
    0 & 0 & 1 & 0 \\
    0 & 0 & 0 & 1
  \end{bmatrix} $ \\
 
 \hline
\end{tabular}
\caption{The Clifford + T gate set}
\label{Clifford table}
\end{table}

\begin{figure}[tbhp]
\flushleft
\small
\[
\Qcircuit @C=0.7em @R=0.5em @!R{
&	&\ctrl{2}		&\qw	&		&		&\qw &\gate{T}	&\qw		&\targ		&\qw			&\ctrl{2}		&\qw		&\ctrl{1}		&\gate{T^{\dag}}	&\qw			&\ctrl{2}		&\targ		&\qw		&\\
&	&\ctrl{1}		&\qw		& = 	&		&\qw		&\gate{T}	&\qw		&\ctrl{-1}		&\targ		&\qw			&\gate{T^{\dag}}	&\targ		&\gate{T^{\dag}}	&\targ		&\qw			&\ctrl{-1}		&\qw		&\\
&	&\targ		&\qw		&		&	&\gate{H}		&\gate{T}	&\qw	&\qw		&\ctrl{-1}		&\targ		&\qw	&\qw		&\gate{T}	&\ctrl{-1}		&\targ		&\gate{H}	&\qw		&
}
\]
\caption{The Toffoli gate and its fault tolerant Clifford + T gate implementation \cite{Maslov}.  This fault tolerant Clifford + T gate implementation of the Toffoli gate has a T-count of $7$. }
\label{mult:FIG6}
\end{figure}
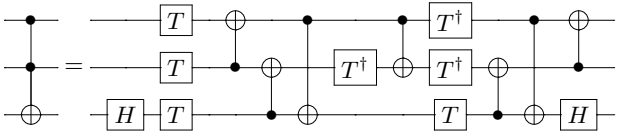

\subsection{Quantum Gates}

Fault tolerant implementation of quantum circuits is gaining the attention of researchers because physical quantum computers are prone to noise errors \cite{Webster} \cite{Zhou} \cite{Polian_DAC}.  Recently, researchers have implemented quantum logic gates and circuits with the fault tolerant \textit{Clifford + T} gate set due to its demonstrated tolerance to noise errors \cite{Miller} \cite{Maslov}.  \textit{The quantum integer multiplication circuit proposed in this work is comprised of Not, Feynman and Toffoli gates exclusively.}  Table \ref{Clifford table} illustrates that the Feynman gate and the Not gate are in the set of gates that make up the Clifford + T gate family.  The Toffoli gate is a 3 input, 3 output logic gate and has the mapping $A,B,C$ to $A,B,A \cdot B \oplus C$. \textit{In this work, we use the fault tolerant implementation of the Toffoli gate designed in \cite{Maslov} shown in figure \ref{mult:FIG6}}.  %The existing fault tolerant Clifford + T gate implementation of the Toffoli gate shown in figure \ref{mult:FIG6} has a T-depth of $3$ and T-count of $7$.}

% include graphic showing all Clifford + T gates

% include graphic showing the fault tolerant implementation of the Toffoli gate figure of 30

\subsection{Evaluation of Quantum Circuit Performance}

Evaluating fault tolerant quantum circuit performance in terms of T-count is of interest to researchers because the fault tolerant implementation costs of the T gate is significantly greater than the fault tolerant implementation costs of the other Clifford + T gates \cite{Mosca2} \cite{Gosset}.  \textit{T-count is the total number of T gates or Hermitian transposes of the T gate in a quantum circuit.}  The Clifford + T implementation of the Toffoli gate illustrated in figure \ref{mult:FIG6} has a T-count of $7$.

%\item \textbf{T-depth:} T-depth of a Clifford + T circuit is the maximum number of T gates or Hermitian transposes of the T gate encountered on any given qubit in the quantum circuit.  The Clifford + T implementation of the Toffoli gate illustrated in figure \ref{mult:FIG6} has a T-depth of $3$.

%\end{itemize}

\subsection{Shift and Add Multiplication Algorithm}

Algorithms for the multiplication of integers in hardware have drawn the interest of researchers.  Researchers have developed many multiplication algorithms such as shift and add, Booth's algorithm and Karatsuba's algorithm.  \textit{In this work, we present a quantum implementation of the shift and add multiplication algorithm optimized for T-count}.

Consider the multiplication of two $n$ bit numbers $a$ and $b$.  At the end of computation, the shift and add multiplication algorithm returns the product $p$ of the multiplication of the two numbers $a$ and $b$.  The steps of the shift and add multiplication algorithm are illustrated for the multiplication of the number $a$ by the number $b$.

\begin{itemize}
\item Step 1: Assign the value $0$ to the product $p$.
\item Step 2: For $i = 0:1:n-1$:

Calculate $p = p + (a \land b_i) \cdot 2^i$
 
\item Step 3: Return product $p$

\end{itemize}

%\cite{Flynn}

\subsection{Related Work}
\label{md_ref_past_stuff}

The design of quantum integer multiplication circuits has received notable attention in the literature.  However, most works 
target reversible computing and suffer from high garbage output costs \cite{Haghparast} \cite{Akbar} \cite{Jamal}.  The works 
proposed in \cite{Lidia}, \cite{Picca}, \cite{Lin} and \cite{Jayashree} are appropriate for quantum computation.  
Papers \cite{Lidia} and \cite{Picca} present quantum multiplication circuits in the quantum Fourier transform (QFT) domain.  
While garbage-less in nature, these circuits have significant Clifford + T gate costs \cite{Kliuchnikov}.  The quantum integer multiplication 
circuits presented in papers \cite{Lin} and \cite{Jayashree} require significantly fewer quantum gates to realize 
and are garbage-less.  The existing quantum integer multiplication circuits are made using Not, Square Root of Not, Feynman 
and Toffoli gates \cite{Lin} \cite{Jayashree} \cite{Babu}.    The Not and Feynman gates are members of the Clifford + T gate 
family \cite{Miller}.  The Square Root of Not and Toffoli gates can be realized with 7 and 15 Clifford + T gates respectively 
\cite{Maslov}.   Further, in a recent work, a quantum integer multiplication circuit design is proposed \cite{Babu}.  The design presented in \cite{Babu} consists of a new quantum AND circuit and quantum full adder circuit.  The quantum AND 
circuit and quantum full adder are created with Feynman gates and square root of not gates.  The multiplication circuit itself 
is based on a two step algorithm: (i) create all partial products with the quantum AND circuit and (ii) combine all partial 
products with the quantum full adder.  To reduce circuit depth, the partial products are realized in parallel.  Further, to 
reduce the depth of the partial product addition, a design methodology based on a partial product addition tree is used.  
While the design in \cite{Babu} is optimized in terms of depth, this design suffers from significant ancillae and garbage 
output costs.  While the integer multiplication circuits presented in papers \cite{Lin}, \cite{Jayashree} and \cite{Babu} are interesting designs, these integer multipliers have significant gate overhead in terms of T-count.  

\subsection{Methodology to Remove Garbage Outputs from Quantum Integer Multiplication Circuit Designs}

In section \ref{md_ref_past_stuff} we presented existing quantum integer multiplication circuits that are garbageless in nature.  However, other recent works (such as the design in by Babu (\cite{Babu})) that show promise in the terms of T-count suffer from significant ancillae and garbage output overhead.  In order to be usable in quantum computing, these designs must be made garbageless in nature.  We can apply the Bennett's garbage removal scheme to make such designs garbageless \cite{Kowada}.

Consider the multiplication of $n$ two bits numbers $a$ and $b$ stored in quantum registers $\ket{A}$ and $\ket{B}$ by a design such as the one in \cite{Babu}.  At the end of computation, the quantum registers $\ket{A}$ and $\ket{B}$ will keep the values $a$ and $b$ respectively.  The product of $a$ and $b$ will appear on a quantum register $\ket{P}$ that is initialized with $\ket{P_i} = 0$ for $0 \leq i \leq 2 \cdot n -1$ .  Further, there will be an additional quantum register $\ket{G}$ that is initialized to $0$.  The quantum register $\ket{G}$, at the end of computation, will hold the garbage outputs.

The Bennett's garbage removal scheme removes the garbage outputs by applying the logical reverse of the original design to the quantum registers $\ket{A}$, $\ket{B}$, $\ket{P}$ and $\ket{G}$.  To preserve the product of $a$ and $b$, the contents of quantum register $\ket{P}$ is copied to another quantum register $\ket{Y}$ that is initialized with $\ket{Y_i} = 0$ for $0 \leq i \leq 2 \cdot n -1$.  Therefore, at the end of computation, the quantum registers $\ket{A}$ and $\ket{B}$ will keep the values $a$ and $b$ respectively.  At the end of computation, the quantum register $\ket{P}$ that originally stored the product will be restored to the value $0$.  Finally, the quantum register $\ket{G}$ that originally stored the garbage outputs will be restored to the value $0$ at the end of computation.

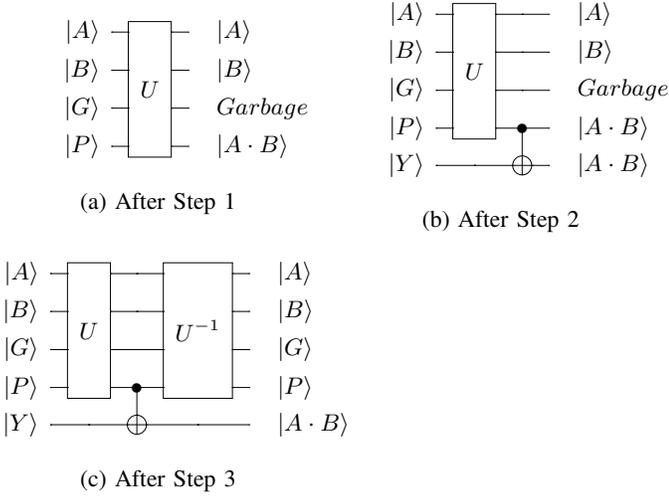
\begin{figure}
	\small
	\flushleft
\begin{subfigure}[thb]{1.5in}
		\flushleft
\[
\Qcircuit @C = .7em @R = .7em {
\lstick{\ket{A}} & \multigate{3}{U} & \qw & \rstick{\ket{A}}\\
\lstick{\ket{B}} & \ghost{U} & \qw & \rstick{\ket{B}}\\
\lstick{\ket{G}} & \ghost{U} & \qw & \rstick{Garbage} \\
\lstick{\ket{P}} & \ghost{U} & \qw & \rstick{\ket{A \cdot B}} \\
}
\]
\caption{After Step 1}
\end{subfigure}  \qquad \begin{subfigure}[thb]{1.5in}

\[
\Qcircuit @C = .7em @R = .7em {
\lstick{\ket{A}} & \multigate{3}{U} & \qw & \qw &\rstick{\ket{A}}\\
\lstick{\ket{B}} & \ghost{U} & \qw & \qw & \rstick{\ket{B}}\\
\lstick{\ket{G}} & \ghost{U} & \qw & \qw & \rstick{Garbage} \\
\lstick{\ket{P}} & \ghost{U} & \ctrl{1}  & \qw & \rstick{\ket{A \cdot B}} \\  
\lstick{\ket{Y}} & \qw &\targ& \qw & \rstick{\ket{A \cdot B}} \\
}
\]
\caption{After Step 2}
\end{subfigure} \\ \begin{subfigure}[thb]{1.5in}

\[
\Qcircuit @C = .7em @R = .7em {
\lstick{\ket{A}} & \multigate{3}{U} & \qw &\multigate{3}{U^{-1}} & \qw &\rstick{\ket{A}}\\
\lstick{\ket{B}} & \ghost{U} & \qw & \ghost{U^{-1}} & \qw & \rstick{\ket{B}}\\
\lstick{\ket{G}} & \ghost{U} & \qw & \ghost{U^{-1}} & \qw & \rstick{\ket{G}} \\
\lstick{\ket{P}} & \ghost{U} & \ctrl{1} & \ghost{U^{-1}} & \qw & \rstick{\ket{P}} \\  
\lstick{\ket{Y}} & \qw &\targ& \qw & \qw & \rstick{\ket{A \cdot B}} \\
}
\]
\caption{After Step 3}
\end{subfigure}
\caption{Generation of garbageless quantum multiplication circuit: steps 1-3.}
\label{figure:bonusmult}
\end{figure}

The steps of the Bennett's garbage removal scheme are explained below.  The methodology is generic and can be used on any quantum integer multiplication circuit that has garbage outputs.  An illustrative example of the methodology for a generic multiplication circuits with garbage outputs is also shown.  Figure \ref{figure:bonusmult} illustrates steps 1 through 3.  The generic multiplication circuit and its logical reverse are labeled in figure \ref{figure:bonusmult} with ``$U$'' and ``$U^{-1}$'' respectively.

\begin{itemize}

\item Step 1: At quantum registers $\ket{A}$, $\ket{B}$, $\ket{P}$ and $\ket{G}$ apply the quantum the multiplication circuit such that the registers $\ket{A}$ and $\ket{B}$ will maintain the same value, location $\ket{P}$ will hold the product and location $\ket{G}$ will contain the garbage outputs.

\item Step 2: For $i = 0:1:2 \cdot n-1$

At locations $\ket{P_i}$ and $\ket{Y_i}$ apply a Feynman gate such that the location $\ket{P_i}$ will maintain the same value while location $\ket{Y_i}$ is transformed to the value in location $\ket{P_i}$.

\item Step 3: At quantum registers $\ket{A}$, $\ket{B}$, $\ket{P}$ and $\ket{G}$ apply the logical reverse quantum the multiplication circuit such that the registers $\ket{A}$ and $\ket{B}$ will maintain the same value, location $\ket{P}$ will be restored to the value $0$ and location $\ket{G}$ will be restored to the value $0$.

\end{itemize}

This methodology, while able to remove garbage outputs from a quantum multiplication designs such as the one in \cite{Babu}, 
does add additional quantum gates and qubit costs.  The methodology will add $2 \cdot n + 1$ ancillae to the design and 
increase the T-count by a factor of at least two.  

%cost impact stuff still pending

%The additional circuitry needed to eliminate the garbage outputs will increase the gate count by a factor of $2$x and add $2 \cdot n + 1$ ancillae to the design.  Further, this quantum integer multiplication circuit design does not take into account fault tolerance.  In order to be able to reliably implement this integer multiplication circuit on quantum hardware, the design must be reimplemented with a fault tolerant quantum gates set such as the Clifford + T gates set.  The fault tolerant Clifford + T gate implementation of the square root of not gate requires 3 T gates and 4 Clifford gates \cite{Maslov}.  Thus, the design in \cite{Babu} will see gate cost increase by a factor of $4.8$x when implemented with the fault tolerant Clifford + T quantum gate set.  To compare against this work, we have added the additional circuitry needed to eliminate the garbage output.  Figure \ref{figure:bonusmult} shows a block diagram of the modified circuit based on the design in \cite{Babu}.  

\section{Design Methodology of Proposed Quantum Conditional Addition Circuit with No Input Carry}
\label{md_adder}

We present the design of the proposed quantum conditional addition (\textit{Ctrl-Add}) circuit with no input carry.  The design has no garbage outputs.  The proposed method improves the T-count of the quantum \textit{Ctrl-Add} circuit compared to existing design approaches which have no garbage outputs.  Consider the conditional addition of two $n$-bit numbers $a_i$ and $b_i$ stored at quantum registers $\ket{A_i}$ and $\ket{B_i}$ respectively (where $0 \leq i \leq n-1$).  The addition of $a_i$ and $b_i$ is conditioned on the value of the 1 bit number $ctrl$ stored at quantum register $\ket{Ctrl}$.  Further, consider that quantum register locations $\ket{A_n}$ and $\ket{A_{n+1}}$ are initialized with $z \in {0,1}$.  At the end of the computation, the quantum register $\ket{B_i}$ will have the value $s_i$ while the quantum register $\ket{A_i}$ keeps the value $a_i$.  The additional quantum register locations $\ket{A_n}$ that initially stored the value $z$ will have the value $z \oplus s_n$ at the end of computation.  Thus, $\ket{A_n}$ will have the value $s_n$ when $z = 0$.  Further, the additional quantum register location $\ket{A_{n+1}}$ that initially stored the value $z$ will have the value $z$ at the end of computation.  Here $s_i$ is the sum bit and is defined as:

\begin{equation}
s_i = \begin{cases}
a_i \oplus b_i \oplus c_i & \qquad \text{if } 0 \leq i \leq n-1 \text{ and } ctrl = 1 \\
c_n & \qquad \text{if } i = n \text{ and } ctrl = 1 \\
b_i & \qquad \text{if } 0 \leq i \leq n-1 \text{ and } ctrl = 0 \\
z & \qquad \text{if } i = n \text{ and } ctrl = 0 \\
\end{cases}
\label{md-equation:1}
\end{equation}

where $c_i$ is the carry bit and is defined as:

\begin{equation}
c_i = \begin{cases}
0 & \, \text{if } i = 0 \\
a_{i-1} \cdot b_{i-1} \oplus b_{i-1} \cdot c_{i-1} \oplus a_{i-1} \cdot c_{i-1} & \, \text{if } 1 \leq i \leq n \\   
\end{cases}
\label{md-equation:2}
\end{equation}

The proposed design methodology of generating the quantum \textit{Ctrl-Add} circuit with no input carry minimizes the garbage outputs by using the strategy first reported in \cite{Boss2}.  When $ctrl = 1$, the carry bits $c_i$ are produced based on the inputs $a_{i-1},b_{i-1}$ and the carry bit $c_{i-1}$ from the previous stage.  All of the generated carry bits $c_i$ are stored on the quantum register $\ket{A_i}$ which initially was used to store $a_i$ for $0 \leq i \leq n-1$.  After the generated carry bits are used in further computation, the quantum register $\ket{A_i}$ is restored to the value $a_i$ while the quantum register $\ket{B_i}$ stores the sum bit $s_i$ for $0 \leq i \leq n-1$.

The proposed design methodology of generating the quantum \textit{Ctrl-Add} circuit with no input carry is explained below.  The proposed methodology is generic and can design a quantum \textit{Ctrl-Add} circuit with no input carry of any size.  The steps involved in the proposed methodology are presented for the conditional addition of two $n$ bit numbers $a_i$ and $b_i$, where $0 \leq i \leq n-1$.  An illustrative example of the generation of a quantum \textit{Ctrl-Add} circuit with no input carry that can perform the conditional addition of two 4 bit numbers $a = a_0 \cdots a_3$ and $b = b_0 \cdots b_3$ is also shown.  Steps 1 through 4 of the methodology are shown in figure \ref{mult:FIG2}.  Steps 5 through 7 of the methodology are shown in figure \ref{mult:FIG3}.

\subsection{Steps of Design Methodology}

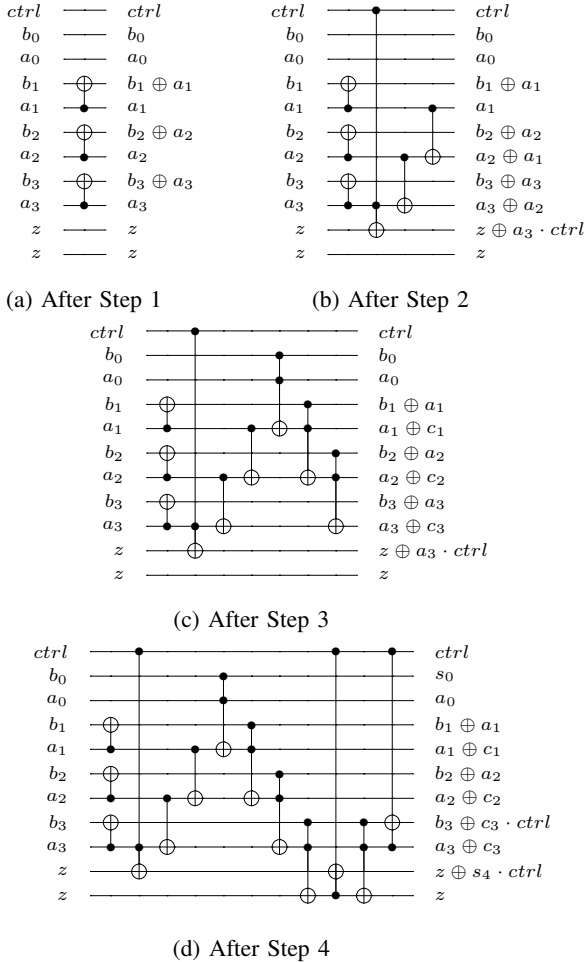
\begin{figure}[tbhp]
\flushleft
\scriptsize
%\resizebox{\textwidth}{!}{
\begin{subfigure}[hb]{1.25in}
\flushleft
\[
\Qcircuit @C=0.7em @R=0.5em @!R{
\lstick{ctrl}&	&\qw		&\qw		&\rstick{ctrl}\\
\lstick{b_0}&	&\qw			&\qw		&\rstick{b_0}\\
\lstick{a_0}&	&\qw			&\qw		&\rstick{a_0}\\
\lstick{b_1}&	&\targ			&\qw		&\rstick{b_1 \oplus a_1}\\
\lstick{a_1}&	&\ctrl{-1}	&\qw		&\rstick{a_1}\\
\lstick{b_2}&	&\targ		&\qw		&\rstick{b_2 \oplus a_2}\\
\lstick{a_2}&	&\ctrl{-1}		&\qw		&\rstick{a_2}\\
\lstick{b_3}&	&\targ		&\qw		&\rstick{b_3 \oplus a_3}\\
\lstick{a_3}&	&\ctrl{-1}		&\qw		&\rstick{a_3}\\
\lstick{z}&	&\qw			&\qw		&\rstick{z}\\
\lstick{z}&	&\qw			&\qw		&\rstick{z}
}
\]
\caption{After Step 1}
\end{subfigure}  \qquad \begin{subfigure}[hb]{1.5in}
	\flushleft
\[
\Qcircuit @C=0.7em @R=0.5em @!R{
\lstick{ctrl}&	&\qw			&\ctrl{9}		&\qw			&\qw			&\qw		&\rstick{ctrl}\\
\lstick{b_0}&		&\qw			&\qw			&\qw			&\qw			&\qw		&\rstick{b_0}\\
\lstick{a_0}&		&\qw			&\qw			&\qw			&\qw			&\qw		&\rstick{a_0}\\
\lstick{b_1}&	&\targ		&\qw			&\qw			&\qw			&\qw		&\rstick{b_1 \oplus a_1}\\
\lstick{a_1}&	&\ctrl{-1}	&\qw			&\qw			&\ctrl{2}		&\qw		&\rstick{a_1}\\
\lstick{b_2}&	&\targ		&\qw			&\qw			&\qw			&\qw		&\rstick{b_2 \oplus a_2}\\
\lstick{a_2}&	&\ctrl{-1}	&\qw			&\ctrl{2}		&\targ		&\qw		&\rstick{a_2 \oplus a_1}\\
\lstick{b_3}&	&\targ		&\qw &\qw			&\qw			&\qw		&\rstick{b_3 \oplus a_3}\\
\lstick{a_3}&	&\ctrl{-1}		&\ctrl{1}		&\targ		&\qw			&\qw		&\rstick{a_3 \oplus a_2}\\
\lstick{z}&	&\qw			&\targ			&\qw			&\qw			&\qw		&\rstick{z \oplus a_3 \cdot ctrl}\\
\lstick{z}&	&\qw			&\qw		&\qw			&\qw			&\qw		&\rstick{z}
}
\]
\caption{After Step 2}
\end{subfigure}	

\begin{subfigure}[hb]{3in}
\flushleft
\[
\Qcircuit @C=0.7em @R=0.5em @!R{
\lstick{ctrl}&	&\qw			&\ctrl{9}		&\qw			&\qw			&\qw			&\qw			&\qw			&\qw		&\rstick{ctrl}\\
\lstick{b_0}&		&\qw			&\qw			&\qw			&\qw			&\ctrl{3}		&\qw			&\qw			&\qw		&\rstick{b_0}\\
\lstick{a_0}&		&\qw			&\qw			&\qw			&\qw			&\ctrl{2}		&\qw			&\qw			&\qw		&\rstick{a_0}\\
\lstick{b_1}&	&\targ		&\qw			&\qw			&\qw			&\qw			&\ctrl{3}		&\qw			&\qw		&\rstick{b_1 \oplus a_1}\\
\lstick{a_1}&	&\ctrl{-1}	&\qw			&\qw			&\ctrl{2}		&\targ		&\ctrl{2}		&\qw			&\qw		&\rstick{a_1 \oplus c_1}\\
\lstick{b_2}&	&\targ		&\qw			&\qw			&\qw			&\qw			&\qw			&\ctrl{3}		&\qw		&\rstick{b_2 \oplus a_2}\\
\lstick{a_2}&	&\ctrl{-1}	&\qw			&\ctrl{2}		&\targ		&\qw			&\targ		&\ctrl{2}		&\qw		&\rstick{a_2 \oplus c_2}\\
\lstick{b_3}&	&\targ		&\qw			&\qw			&\qw			&\qw			&\qw			&\qw			&\qw		&\rstick{b_3 \oplus a_3}\\
\lstick{a_3}&	&\ctrl{-1}		&\ctrl{1}		&\targ		&\qw			&\qw			&\qw			&\targ		&\qw		&\rstick{a_3 \oplus c_3}\\
\lstick{z}&	&\qw			&\targ			&\qw			&\qw			&\qw			&\qw			&\qw			&\qw		&\rstick{z \oplus a_3 \cdot ctrl}\\
\lstick{z}&	&\qw			&\qw		&\qw			&\qw			&\qw			&\qw			&\qw			&\qw		&\rstick{z}
}
\]
\caption{After Step 3}
\end{subfigure}

\begin{subfigure}[hb]{3in}
\flushleft
\[
\Qcircuit @C=0.7em @R=0.5em @!R{
\lstick{ctrl}&	&\qw			&\ctrl{9}		&\qw			&\qw			&\qw			&\qw			&\qw			&\qw			&\ctrl{10}		&\qw			&\ctrl{7}		&\qw		&\rstick{ctrl}\\
\lstick{b_0}&	&\qw			&\qw			&\qw			&\qw			&\ctrl{3}		&\qw			&\qw			&\qw			&\qw			&\qw			&\qw			&\qw		&\rstick{s_0}\\
\lstick{a_0}&	&\qw			&\qw			&\qw			&\qw			&\ctrl{2}		&\qw			&\qw			&\qw			&\qw			&\qw			&\qw			&\qw		&\rstick{a_0}\\
\lstick{b_1}&	&\targ		&\qw			&\qw			&\qw			&\qw			&\ctrl{3}		&\qw			&\qw			&\qw			&\qw			&\qw			&\qw		&\rstick{b_1 \oplus a_1}\\
\lstick{a_1}&	&\ctrl{-1}	&\qw			&\qw			&\ctrl{2}		&\targ		&\ctrl{2}		&\qw			&\qw			&\qw			&\qw			&\qw			&\qw		&\rstick{a_1 \oplus c_1}\\
\lstick{b_2}&	&\targ		&\qw			&\qw			&\qw			&\qw			&\qw			&\ctrl{3}		&\qw			&\qw			&\qw			&\qw			&\qw		&\rstick{b_2 \oplus a_2}\\
\lstick{a_2}&	&\ctrl{-1}		&\qw			&\ctrl{2}		&\targ		&\qw			&\targ		&\ctrl{2}		&\qw			&\qw			&\qw			&\qw			&\qw		&\rstick{a_2 \oplus c_2}\\
\lstick{b_3}&	&\targ		&\qw			&\qw			&\qw			&\qw			&\qw			&\qw			&\ctrl{2}		&\qw			&\ctrl{2}		&\targ		&\qw		&\rstick{b_3 \oplus c_3 \cdot ctrl}\\
\lstick{a_3}&	&\ctrl{-1}		&\ctrl{1}		&\targ		&\qw			&\qw			&\qw			&\targ		&\ctrl{2}		&\qw			&\ctrl{2}		&\ctrl{-1}		&\qw		&\rstick{a_3 \oplus c_3}\\
\lstick{z}&		&\qw			&\targ			&\qw			&\qw			&\qw			&\qw			&\qw			&\qw		&\targ			&\qw		&\qw			&\qw		&\rstick{z \oplus s_4 \cdot ctrl}\\
\lstick{z}&		&\qw			&\qw		&\qw			&\qw			&\qw			&\qw			&\qw			&\targ			&\ctrl{-1}	&\targ			&\qw			&\qw		&\rstick{z}
}
\]
\caption{After Step 4}
\end{subfigure}

\caption{Generation of a 4-qubit quantum \textit{Ctrl-Add} circuit with no input carry: steps 1-4.}
\label{mult:FIG2}
\end{figure}

\begin{itemize}

\item Step 1: For $i = 1:1:n-1$

At each pair of quantum register locations $\ket{A_i}$ and $\ket{B_i}$ apply a Feynman gate such that the location $\ket{A_i}$ will maintain the same value, while location $\ket{B_i}$ is transformed to $\ket{A_i \oplus B_i}$.  %Step 1 is shown for a quantum \textit{Ctrl-Add} circuit with no input carry that can perform the conditional addition of two 4 bit numbers in figure \ref{mult:FIG2}.

\item Step 2: Step 2 has the following two sub-steps:

\begin{itemize}

\item Step 1: At quantum register locations $\ket{Ctrl},\ket{A_{n-1}}$ and $\ket{A_n}$ apply a Toffoli gate such that the locations $\ket{Ctrl}, \ket{A_{n-1}}$ and $\ket{A_n}$ are passed to the inputs $A,B,C$ respectively, of the Toffoli gate.

\item Step 2: For $i = n-2:-1:1$

At each pair of quantum register locations $\ket{A_i}$ and $\ket{A_{i+1}}$ apply a Feynman gate such that the location $\ket{A_i}$ will maintain the same value while the location $\ket{A_{i+1}}$ is transformed to $\ket{A_i \oplus A_{i+1}}$. 

\end{itemize}

%Step 2 is shown for a quantum \textit{Ctrl-Add} circuit with no input carry that can perform the conditional addition of two 4 bit numbers in figure \ref{mult:FIG2}.

\item Step 3: For $i = 0:1:n-2$

At quantum register locations $\ket{B_i}, \ket{A_i}$ and $\ket{A_{i+1}}$ apply a Toffoli gate such that $\ket{B_i}$ and $\ket{A_i}$ and $\ket{A_{i+1}}$ are passed to the inputs $A,B,C$ respectively, of the Toffoli gate.  %Step 3 is shown for a quantum \textit{Ctrl-Add} circuit with no input carry that can perform the conditional addition of two 4 bit numbers in figure \ref{mult:FIG2}.

\item Step 4: Step 4 has the following four sub-steps: 

\begin{itemize}

\item Step 1: At quantum register locations $\ket{B_{n-1}},\ket{A_{n-1}}$ and $\ket{A_{n+1}}$ apply a Toffoli gate such that the locations $\ket{B_{n-1}},\ket{A_{n-1}}$ and $\ket{A_{n+1}}$ are passed to the inputs $A,B,C$ respectively, of the Toffoli gate. 

\item Step 2: At quantum register locations $\ket{Ctrl},\ket{A_{n+1}}$ and $\ket{A_n}$ apply a Toffoli gate such that the locations $\ket{Ctrl},\ket{A_{n+1}}$ and $\ket{A_n}$ are passed to the inputs $A,B,C$ respectively, of the Toffoli gate.

\item Step 3: At quantum register locations $\ket{B_{n-1}},\ket{A_{n-1}}$ and $\ket{A_{n+1}}$ apply a Toffoli gate such that the locations $\ket{B_{n-1}},\ket{A_{n-1}}$ and $\ket{A_{n+1}}$ are passed to the inputs $A,B,C$ respectively, of the Toffoli gate. 

\item Step 4: At quantum register locations $\ket{Ctrl},\ket{A_{n-1}}$ and $\ket{B_{n-1}}$ apply a Toffoli gate such that the locations $\ket{Ctrl},\ket{A_{n-1}}$ and $\ket{B_{n-1}}$ are passed to the inputs $A,B,C$ respectively, of the Toffoli gate. 

\end{itemize}

%Step 4 is shown for a quantum \textit{Ctrl-Add} circuit with no input carry that can perform the conditional addition of two 4 bit numbers in figure \ref{mult:FIG2}.

\item Step 5: For $i = n-2:-1:0$. Step 5 has the following two sub-steps:

\begin{itemize}

\item Step 1: At quantum register locations $\ket{B_i},\ket{A_i}$ and $\ket{A_{i+1}}$ apply a Toffoli gate such that the locations $\ket{B_i},\ket{A_i}$ and $\ket{A_{i+1}}$ are passed to the inputs $A,B,C$ respectively, of the Toffoli gate. 

\item Step 2: At quantum register locations $\ket{Ctrl},\ket{A_i}$ and $\ket{B_i}$ apply a Toffoli gate such that the locations $\ket{Ctrl},\ket{A_i}$ and $\ket{B_i}$ are passed to the inputs $A,B,C$ respectively, of the Toffoli gate. 
 
\end{itemize}

%Step 5 is shown for a quantum \textit{Ctrl-Add} circuit with no input carry that can perform the conditional addition of two 4 bit numbers in figure \ref{mult:FIG3}.

\item Step 6: For $i = 1:1:n-2$

At each pair of quantum register locations $\ket{A_i}$ and $\ket{A_{i+1}}$ apply a Feynman gate such that the location $\ket{A_i}$ will maintain the same value while the location $\ket{A_{i+1}}$ is transformed to $\ket{A_i \oplus A_{i+1}}$.  %Step 6 is shown for a quantum \textit{Ctrl-Add} circuit with no input carry that can perform the conditional addition of two 4 bit numbers in figure \ref{mult:FIG3}.

\item Step 7: For $i = 1:1:n-1$

At each pair of quantum register locations $\ket{A_i}$ and $\ket{B_i}$ apply a Feynman gate such that the location $\ket{A_i}$ will maintain the same value, while location $\ket{B_i}$ is transformed to $\ket{A_i \oplus B_i}$.  %Step 7 is shown for a quantum \textit{Ctrl-Add} circuit with no input carry that can perform the conditional addition of two 4 bit numbers in figure \ref{mult:FIG3}.

\end{itemize} 

\begin{figure}[t!bhp]
\flushleft
\scriptsize
%\resizebox{\textwidth}{!}{
\begin{subfigure}[hb]{3in}
\flushleft
\[
\Qcircuit @C=0.3em @R=0.5em @!R{
\lstick{ctrl}&	&\qw			&\ctrl{10}		&\qw			&\qw			&\qw			&\qw			&\qw			&\qw			&\ctrl{10}		&\qw			&\ctrl{7}		&\qw			&\ctrl{5}		&\qw			&\ctrl{3}		&\qw			&\ctrl{1}		&\qw		&\rstick{ctrl}  \\
\lstick{b_0}&		&\qw			&\qw			&\qw			&\qw			&\ctrl{3}		&\qw			&\qw			&\qw			&\qw			&\qw			&\qw			&\qw			&\qw			&\qw			&\qw			&\ctrl{3}		&\targ		&\qw		&\rstick{s_0}  \\
\lstick{a_0}&	&\qw			&\qw			&\qw			&\qw			&\ctrl{2}		&\qw			&\qw			&\qw			&\qw			&\qw			&\qw			&\qw			&\qw			&\qw			&\qw			&\ctrl{2}		&\ctrl{-1}		&\qw		&\rstick{a_0}  \\
\lstick{b_1}&	&\targ		&\qw			&\qw			&\qw			&\qw			&\ctrl{3}		&\qw			&\qw			&\qw			&\qw			&\qw			&\qw			&\qw			&\ctrl{3}		&\targ		&\qw			&\qw			&\qw		&\rstick{b_1 \oplus c_1 \cdot ctrl}  \\
\lstick{a_1}&	&\ctrl{-1}		&\qw			&\qw			&\ctrl{2}		&\targ		&\ctrl{2}		&\qw			&\qw			&\qw			&\qw			&\qw			&\qw			&\qw			&\ctrl{2}		&\ctrl{-1}		&\targ		&\qw			&\qw		&\rstick{a_1}  \\
\lstick{b_2}&	&\targ		&\qw			&\qw			&\qw			&\qw			&\qw			&\ctrl{3}		&\qw			&\qw			&\qw			&\qw			&\ctrl{3}		&\targ		&\qw			&\qw			&\qw			&\qw			&\qw		&\rstick{b_2 \oplus c_2 \cdot ctrl}  \\
\lstick{a_2}&	&\ctrl{-1}		&\qw			&\ctrl{2}		&\targ		&\qw			&\targ		&\ctrl{2}		&\qw			&\qw			&\qw			&\qw			&\ctrl{2}		&\ctrl{-1}		&\targ		&\qw			&\qw			&\qw			&\qw		&\rstick{a_2 \oplus a_1} \\
\lstick{b_3}&	&\targ		&\qw			&\qw			&\qw			&\qw			&\qw			&\qw			&\ctrl{2}		&\qw			&\ctrl{2}		&\targ		&\qw			&\qw			&\qw			&\qw			&\qw			&\qw			&\qw		&\rstick{b_3 \oplus c_3 \cdot ctrl}  \\
\lstick{a_3}&	&\ctrl{-1}		&\ctrl{2}		&\targ		&\qw			
&\qw			&\qw			&\targ		&\ctrl{2}		&\qw			
&\ctrl{2}		&\ctrl{-1}		&\targ		&\qw			&\qw			
&\qw			&\qw			&\qw			&\qw		&\rstick{a_3 
\oplus a_2}  \\
\lstick{z}&	&\qw			&\qw			&\qw			&\qw			
&\qw			&\qw			&\qw			&\qw		&\targ		
&\qw		&\qw			&\qw			&\qw			&\qw			
&\qw			&\qw			&\qw			&\qw		&\rstick{z \oplus 
s_4 \cdot ctrl}  \\
\lstick{z}&	&\qw			&\targ		&\qw			&\qw			
&\qw			&\qw			&\qw			&\targ			
&\ctrl{-1}		&\targ			&\qw			&\qw			
&\qw			&\qw			&\qw			&\qw			
&\qw			&\qw		&\rstick{z}  
}
\]
\caption{After Step 5}
\end{subfigure}

\begin{subfigure}[hb]{3in}
\flushleft
\[
\Qcircuit @C=0.3em @R=0.5em @!R{
\lstick{ctrl}&	&\qw			&\ctrl{10}		&\qw			&\qw			&\qw			&\qw			&\qw			&\qw			&\ctrl{10}		&\qw			&\ctrl{7}		&\qw			&\ctrl{5}		&\qw			&\ctrl{3}		&\qw			&\ctrl{1}		&\qw			&\qw		&\rstick{ctrl} \\
\lstick{b_0}&	&\qw			&\qw			&\qw			&\qw			&\ctrl{3}		&\qw			&\qw			&\qw			&\qw			&\qw			&\qw			&\qw			&\qw			&\qw			&\qw			&\ctrl{3}		&\targ		&\qw			&\qw		&\rstick{s_0} \\
\lstick{a_0}&	&\qw			&\qw			&\qw			&\qw			&\ctrl{2}		&\qw			&\qw			&\qw			&\qw			&\qw			&\qw			&\qw			&\qw			&\qw			&\qw			&\ctrl{2}		&\ctrl{-1}		&\qw			&\qw		&\rstick{a_0}  \\
\lstick{b_1}&	&\targ		&\qw			&\qw			&\qw			&\qw			&\ctrl{3}		&\qw			&\qw			&\qw			&\qw			&\qw			&\qw			&\qw			&\ctrl{3}		&\targ		&\qw			&\qw			&\qw			&\qw		&\rstick{b_1 \oplus c_1 \cdot ctrl}  \\
\lstick{a_1}&	&\ctrl{-1}		&\qw			&\qw			&\ctrl{2}		&\targ		&\ctrl{2}		&\qw			&\qw			&\qw			&\qw			&\qw			&\qw			&\qw			&\ctrl{2}		&\ctrl{-1}		&\targ				&\ctrl{2}		&\qw			&\qw		&\rstick{a_1}  \\
\lstick{b_2}&	&\targ		&\qw			&\qw			&\qw			&\qw			&\qw			&\ctrl{3}		&\qw			&\qw			&\qw			&\qw			&\ctrl{3}		&\targ		&\qw			&\qw			&\qw			&\qw			&\qw			&\qw		&\rstick{b_2 \oplus c_2 \cdot ctrl}  \\
\lstick{a_2}&	&\ctrl{-1}			&\qw			&\ctrl{2}		&\targ		&\qw			&\targ		&\ctrl{2}		&\qw			&\qw			&\qw			&\qw			&\ctrl{2}		&\ctrl{-1}		&\targ		&\qw			&\qw					&\targ		&\ctrl{2}		&\qw		&\rstick{a_2}  \\
\lstick{b_3}&	&\targ		&\qw			&\qw			&\qw			&\qw			&\qw			&\qw			&\ctrl{2}		&\qw			&\ctrl{2}		&\targ		&\qw			&\qw			&\qw			&\qw			&\qw			&\qw			&\qw			&\qw		&\rstick{b_3 \oplus c_3 \cdot ctrl}  \\
\lstick{a_3}&	&\ctrl{-1}		&\ctrl{2}		&\targ		&\qw			
&\qw			&\qw			&\targ		&\ctrl{2}		&\qw			
&\ctrl{2}		&\ctrl{-1}		&\targ		&\qw			&\qw			
&\qw			&\qw			&\qw			&\targ		&\qw		
&\rstick{a_3}  \\
\lstick{z}&	&\qw			&\qw			&\qw			&\qw			
&\qw			&\qw			&\qw				&\qw		
&\targ		&\qw		&\qw			&\qw			&\qw			
&\qw			&\qw			&\qw			&\qw			
&\qw			&\qw		&\rstick{z \oplus s_4 \cdot ctrl}  \\
\lstick{z}&	&\qw			&\targ		&\qw			&\qw			
&\qw			&\qw			&\qw			&\targ			
&\ctrl{-1} 		&\targ			&\qw			&\qw			
&\qw			&\qw			&\qw			&\qw			
&\qw			&\qw			&\qw		&\rstick{z}  
}
\]
\caption{After Step 6}
\end{subfigure}

\begin{subfigure}[hb]{3in}
\flushleft
\[
\Qcircuit @C=0.3em @R=0.5em @!R{
\lstick{ctrl}&	&\qw			&\ctrl{10}		&\qw			&\qw			&\qw			&\qw			&\qw			&\qw			&\ctrl{10}		&\qw			&\ctrl{7}		&\qw			&\ctrl{5}		&\qw			&\ctrl{3}		&\qw			&\ctrl{1}			&\qw			&\qw			&\qw			&\rstick{ctrl}  \\
\lstick{b_0}&		&\qw			&\qw			&\qw			&\qw			&\ctrl{3}		&\qw			&\qw			&\qw			&\qw			&\qw			&\qw			&\qw			&\qw			&\qw			&\qw			&\ctrl{3}		&\targ			&\qw			&\qw			&\qw			&\rstick{s_0} \\
\lstick{a_0}&		&\qw			&\qw			&\qw			&\qw			&\ctrl{2}		&\qw			&\qw			&\qw			&\qw			&\qw			&\qw			&\qw			&\qw			&\qw			&\qw			&\ctrl{2}		&\ctrl{-1}		&\qw			&\qw			&\qw			&\rstick{a_0} \\
\lstick{b_1}&	&\targ			&\qw			&\qw			&\qw			&\qw			&\ctrl{3}		&\qw			&\qw			&\qw			&\qw			&\qw			&\qw			&\qw			&\ctrl{3}		&\targ		&\qw			&\qw			&\qw				&\targ		&\qw				&\rstick{s_1} \\
\lstick{a_1}&	&\ctrl{-1}		&\qw			&\qw			&\ctrl{2}		&\targ		&\ctrl{2}		&\qw			&\qw			&\qw			&\qw			&\qw			&\qw			&\qw			&\ctrl{2}		&\ctrl{-1}		&\targ				&\ctrl{2}		&\qw			&\ctrl{-1}		&\qw			&\rstick{a_1} \\
\lstick{b_2}&	&\targ		&\qw			&\qw			&\qw			&\qw			&\qw			&\ctrl{3}		&\qw			&\qw			&\qw			&\qw			&\ctrl{3}		&\targ		&\qw			&\qw			&\qw			&\qw			&\qw						&\targ &\qw		&\rstick{s_2} \\
\lstick{a_2}&	&\ctrl{-1}			&\qw			&\ctrl{2}		&\targ		&\qw			&\targ		&\ctrl{2}		&\qw			&\qw			&\qw			&\qw			&\ctrl{2}		&\ctrl{-1}		&\targ		&\qw			&\qw					&\targ		&\ctrl{2}				&\ctrl{-1}	&\qw	&\rstick{a_2} \\
\lstick{b_3}&	&\targ		&\qw			&\qw			&\qw			&\qw			&\qw			&\qw			&\ctrl{2}		&\qw			&\ctrl{2}		&\targ		&\qw			&\qw			&\qw			&\qw			&\qw			&\qw			&\qw		&\targ		&\qw 	&\rstick{s_3} \\
\lstick{a_3}&	&\ctrl{-1}		&\ctrl{2}		&\targ		&\qw			
&\qw			&\qw			&\targ		&\ctrl{2}		&\qw			
&\ctrl{2}		&\ctrl{-1}		&\targ		&\qw			&\qw			
&\qw			&\qw			&\qw				&\targ			
&\ctrl{-1}	&\qw	&\rstick{a_3} \\
\lstick{z}&	&\qw			&\qw			&\qw			&\qw			
&\qw			&\qw			&\qw			&\qw		&\targ		
&\qw		&\qw			&\qw			&\qw			
&\qw			&\qw			&\qw			&\qw			
&\qw			&\qw			&\qw			&\rstick{z \oplus s_4 \cdot 
ctrl} \\
\lstick{z}&	&\qw			&\targ		&\qw			&\qw			
&\qw			&\qw			&\qw			&\targ			
&\ctrl{-1} 		&\targ			&\qw			&\qw			
&\qw			&\qw			&\qw			&\qw			
&\qw			&\qw			&\qw			&\qw			&\rstick{z}
}
\]
\caption{After Step 7}
\end{subfigure}

\caption{Generation of a 4-qubit quantum \textit{Ctrl-Add} circuit with no input carry: steps 5-7.}
\label{mult:FIG3}

\end{figure}
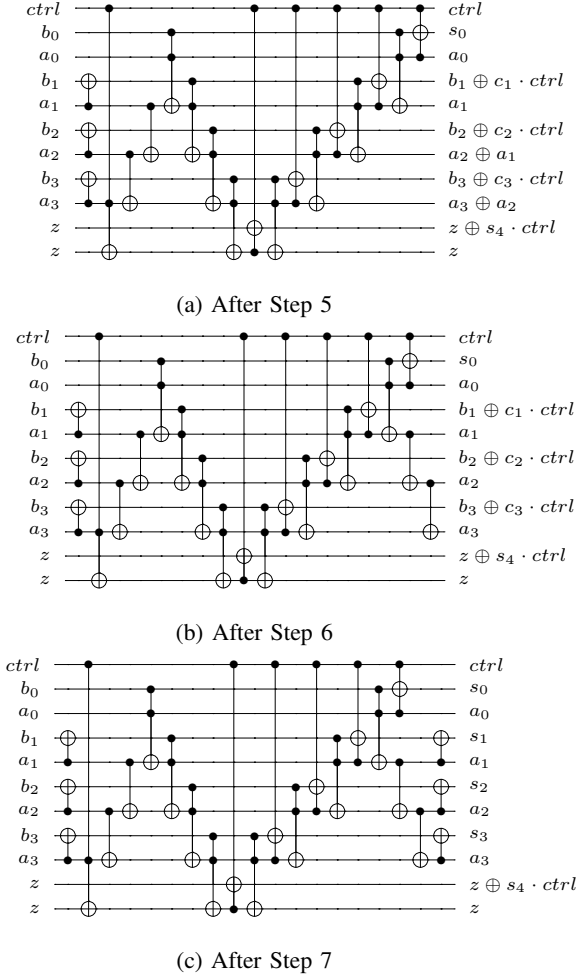

\textbf{Theorem:} \textit{Let $a$ and $b$ be two $n$ bit binary numbers 
represented as $a_i$ and $b_i$, $ctrl \in {0,1}$ be a 1-bit input and $z \in 
{0,1}$ is another 1-bit input, where $0 \leq i \leq n-1$, then the proposed 
design methodology results in a quantum \textit{Ctrl-Add} circuit with no 
input carry that functions correctly.  The proposed design methodology designs 
a $n$-bit \textit{Ctrl-Add} circuit that produces the sum output $s_i$ at the 
quantum register where $b_i$ is stored, while it restores the quantum register 
where $a_i$ is initially stored to the value $a_i$ for $0 \leq i \leq n-1$.  
Further, the additional quantum register locations $\ket{A_{n}}$ and 
$\ket{A_{n+1}}$ where $z$ is initially stored will have the values $z \oplus 
s_n$ and $z$ respectively.}

\textbf{Proof:} The proposed design methodology will make the following 
changes to the inputs.  An example of the transformation of the inputs states after Steps 1 through 4 is 
illustrated for a 4-bit quantum \textit{Ctrl-Add} circuit with no input carry in figure \ref{mult:FIG2}.  An example of the transformation of the inputs states after Steps 5 through 7 is 
illustrated for a 4-bit quantum \textit{Ctrl-Add} circuit with no input carry in figure \ref{mult:FIG3}

\begin{itemize}

\item Step 1:
Step 1 of the proposed design methodology transforms the input states to: 

\end{itemize}

\begin{equation}
\ket{b_0} \ket{a_0} \left( \bigoplus_{i = 1}^{n-1} \ket{b_i \oplus a_i} 
\ket{a_i} \right) \ket{z} \ket{z}
\label{md-equation:3}
\end{equation}

\begin{itemize}

%\item[] An example of the transformation of the inputs states after Step 1 is illustrated for a 4-bit quantum \textit{Ctrl-Add} circuit with no input carry in figure \ref{mult:FIG2}.

\item Step 2:
Step 2 of the proposed design methodology transforms the input states $\ket{a_i}$ and $\ket{b_i}$ for $0 \leq i \leq n-1$ to: 

\end{itemize}

\begin{equation}
\ket{b_0} \ket{a_0} \ket{b_1 \oplus a_1} \ket{a_1} \left( \bigoplus_{i = 
2}^{n-1} \ket{b_i \oplus a_i} \ket{a_i \oplus a_{i-1}} \right) 
\label{md-equation:4-a}
\end{equation}

\begin{itemize}

\item[] Step 2 of the proposed design methodology transforms the remaining input states $\ket{a_n}, \ket{a_{n+1}}$ to:

\end{itemize}

\begin{equation}
\ket{z \oplus a_{n-1} \cdot ctrl} \ket{z}
\label{md-equation:4-b}
\end{equation}

\begin{itemize}

%\item[] An example of the transformation of the input states after Step 2 is illustrated for a 4-bit quantum \textit{Ctrl-Add} circuit in figure \ref{mult:FIG2}.

\item Step 3:
Step 3 has $n-1$ Toffoli gates.  The first Toffoli gate takes $b_0$, $a_0$ and 
$a_1$ as inputs and produces the output as $b_0$, $a_0$ and $a_1 \oplus c_1$ 
where $c_1$ represents the generated output carry after the addition of $b_0$ 
and $a_0$.  The remaining $n-2$ Toffoli gates take $a_i \oplus b_i$, $a_i 
\oplus c_i$ and $a_i \oplus a_{i+1}$ as inputs and produces the outputs as 
$a_i \oplus b_i$, $a_i \oplus c_i$ and $a_{i+1} \oplus c_{i+1}$.  Thus, after 
Step 3, input states are transformed to:

%\end{itemize}

\begin{equation}
\ket{b_0} \ket{a_0} \left( \bigoplus_{i = 1}^{n-1} \ket{b_i \oplus a_i} 
\ket{a_i \oplus c_i} \right) \ket{z \oplus a_{n-1} \cdot ctrl} \ket{z}
\label{md-equation:5}
\end{equation}

%\begin{itemize}

%An example of the transformation the input states after Step 3 is illustrated for a 4-bit quantum \textit{Ctrl-Add} circuit in figure \ref{mult:FIG2}.

\item Step 4:

Step 4 has four Toffoli gates.  The first Toffoli gate takes $a_{n-1} \oplus 
b_{n-1}$, $c_{n-1} \oplus a_{n-1}$ and $z$ as inputs to produce the outputs as 
$a_{n-1} \oplus b_{n-1}$, $c_{n-1} \oplus a_{n-1}$ and $z \oplus 
(\overline{a}bc + a\overline{bc})$ where $\overline{a,b}$ and $\overline{c}$ 
are the compliments of $a,b$ and $c$.  The third output of the Toffoli gate is 
$z \oplus (\overline{a}bc + a\overline{bc})$ because the gate realizes $A 
\cdot B \oplus C$ where $A,B$ and $C$ are inputs to the Toffoli gate.  Thus, 
the Toffoli gate will have the third input as $z \oplus (a_{n-1} \oplus 
b_{n-1}) \cdot (c_{n-1} \oplus a_{n-1}) = z \oplus (\overline{a}bc + 
a\overline{bc})$.  The second Toffoli gate takes $ctrl$, $z \oplus 
(\overline{a}bc + a\overline{bc})$ and $z \oplus a_{n-1} \cdot ctrl$ is inputs 
to produce the outputs as $ctrl$, $z \oplus (\overline{a}bc + a\overline{bc})$ 
and $z \oplus c_n \cdot ctrl$.  Note that $z \oplus c_n \cdot ctrl = z \oplus 
s_n$.  The third Toffoli gate takes $a_{n-1} \oplus b_{n-1}$, $c_{n-1} \oplus 
a_{n-1}$ and $z \oplus (\overline{a}bc + a\overline{bc})$ as inputs to produce 
the outputs as $a_{n-1} \oplus b_{n-1}$, $c_{n-1} \oplus a_{n-1}$ and $z$.  
The fourth Toffoli gates takes the inputs $ctrl, a_{n-1} \oplus c_{n-1}$ and 
$a_{n-1} \oplus b_{n-1}$ to produce the outputs as $ctrl, a_{n-1} \oplus 
c_{n-1}$ and $b_{n-1} \oplus c_{n-1} \cdot ctrl$.  Thus, Step 4 transforms the 
input states $\ket{a_i}$ and $\ket{b_i}$ for $0 \leq i \leq n-2$ to:

\end{itemize}

%\end{multicols}

%!!!!!!!!!!!!!!!!!!!!!!!!!!!!!!!!!!!!!!!!!!!!!!!!!!!!!!!!!!!!!!!!!!!!
%\begin{widetext}
%\begin{figure*}[t!hbp]
\begin{equation}
\ket{s_0} \ket{a_0} \left( \bigoplus_{i = 1}^{n-2} \ket{b_i \oplus a_i} 
\ket{a_i \oplus c_i} \right) 
\label{md-equation:6-a}
\end{equation}

\begin{itemize}

\item[] Step 4 transforms the remaining input states $\ket{a_{n-1}}, \ket{b_{n-1}}, \ket{a_{n}}, \ket{a_{n+1}}$ to:

\end{itemize}

\begin{equation}
\ket{b_{n-1} \oplus c_{n-1} \cdot ctrl} 
\ket{a_{n-1} \oplus c_{n-i}} \ket{z \oplus s_{n} } \ket{z}
\label{md-equation:6-b}
\end{equation}
%\end{figure*}
%\end{widetext}

%\begin{multicols}{2}

\begin{itemize}

%\item[] An example of the transformation of the input states after Step 4 is illustrated for a 4-bit quantum \textit{Ctrl-Add} circuit in figure \ref{mult:FIG2}.

%\item[]
\item[]
\item[]

\item Step 5:

Step 5 has $2 \cdot n-2$ Toffoli gates.  The transformations performed by the 
first $2 \cdot n-4$ Toffoli gates are illustrated below:

For $i = n-1:1:2 $:
Toffoli gate $2 \cdot i$ takes $b_{i-1} \oplus a_{i-1}$, $c_{i-1} \oplus 
a_{i-1}$ and $c_{i} \oplus a_{i}$ as inputs to produce the outputs as $b_{i-1} 
\oplus a_{i-1}$, $c_{i-1} \oplus a_{i-1}$ and $a_{i} \oplus a_{i-1}$. Toffoli 
gate $2 \cdot i -1$ takes $ctrl$, $c_{i-1} \oplus a_{i-1}$ and $b_{i-1} \oplus 
a_{i-1}$ as inputs to produce the outputs as $ctrl$, $c_{i-1} \oplus a_{i-1}$ 
and $b_{i-1} \oplus c_{i-1} \cdot ctrl$.

Toffoli gate $2 \cdot n - 3$ takes $b_0$, $a_0$ and $c_1 \oplus a_1$ as inputs 
to produce the outputs as $b_0$, $a_0$ and $a_1$.  Toffoli gate $2 \cdot n - 
2$ takes $ctrl$, $a_0$ and $b_0$ as inputs to produce the outputs as $ctrl$, 
$a_0$ and $s_0$.  Thus, Step 5 transforms the input states $\ket{a_i}$ and $\ket{b_i}$ for $i = 0$ and $i = 1$ to:

%\onecolumngrid
\end{itemize}

\begin{equation}
\ket{s_0} \ket{a_0} \ket{b_1 \oplus c_1 \cdot ctrl} \ket{a_1} 
\label{md-equation:7-a}
\end{equation}

\begin{itemize}
\item[] Step 5 transforms the input states $\ket{a_j}$ and $\ket{b_i}$ for $2 \leq j \leq n+1$ and $2 \leq i \leq n-1$ to:

\end{itemize}
\begin{equation}
\left( \bigoplus_{i = 2}^{n-1} \ket{b_i \oplus c_i \cdot ctrl} \ket{a_i \oplus a_{i-1}} \right) \ket{z \oplus s_{n} } \ket{z}
\label{md-equation:7-b}
\end{equation}

\begin{itemize}

%\twocolumngrid

% \item[] An example of the transformation of the input states after Step 5 is illustrated for a 4-bit quantum \textit{Ctrl-Add} circuit in figure \ref{mult:FIG3}.

\item Step 6:

Step 6 of the proposed design methodology transforms the input states to:

\begin{equation}
\ket{s_0} \ket{a_0} \left( \bigoplus_{i = 1}^{n-1} \ket{b_i \oplus c_i \cdot 
ctrl} \ket{a_i} \right) \ket{z \oplus s_{n} } \ket{z}
\label{md-equation:8}
\end{equation}

% An example of the transformation of the input states after Step 6 is illustrated for a 4-bit quantum \textit{Ctrl-Add} circuit in figure \ref{mult:FIG3}.

\item Step 7:

Step 7 of the proposed design methodology transforms that the input states to:

\begin{equation}
\left( \bigoplus_{i = 0}^{n-1} \ket{s_i} \ket{a_i} \right) \ket{z \oplus s_{n} 
} \ket{z}
\label{md-equation:9}
\end{equation}

% An example of the transformation of the input states after Step 7 is illustrated for a 4-bit quantum \textit{Ctrl-Add} circuit in figure \ref{mult:FIG3}.

\end{itemize}

Thus, the proposed design methodology transforms the quantum register where 
$b_i$ is initially stored to the sum $s_i$, while the quantum register where 
$a_i$ is originally stored will be restored to the value $a_i$.  The 
additional quantum register locations $\ket{A_n}$ and $\ket{A_{n+1}}$ where 
$z$ is initially stored will have $z \oplus s_{n}$ and $z$ respectively.  
Hence, the proposed design methodology generates a quantum \textit{Ctrl-Add} 
circuit with no input carry that is functionally correct.

\subsection{Cost Analysis}

\begin{table}[tbhp]
\caption{Comparison of quantum \textit{Ctrl-Add} circuits}
\label{md-table:1}
\centering
%\resizebox{\columnwidth}{!}{
\begin{tabular}{ccccc}
%\multicolumn{5}{c}{Comparison of quantum \textit{Ctrl-Add} circuits}		\\ 
\midrule										
	&		& 		1		& 		2		&		Proposed		\\ 
	\cmidrule{1-1} \cmidrule{3-5}
T-count	&		&	$	56 \cdot n	$	&	$	28 \cdot n + 7	$	&	$	21 
\cdot n + 14	$	\\
 & & & & \\
% T-depth	&		&	$	\begin{cases} 14 \text{ If } n < 4 \\ 4 \cdot n \text{ If } n \geq 4 \end{cases}	$	&	$	7 \cdot n + 3	$	&	$	\begin{cases}  14 \text{ If } n \leq 5 \\ 2 \cdot n + 4 \text{ If } n > 5 \end{cases}	$	\\ 
qubits 	&		&	$2 \cdot n + 1$			&		$2 \cdot n + 1$		&		$2 \cdot n + 1$		\\
%garbage & & 0 & 0 & 0 \\
ancillae & & 2 & 2  & 2 \\

 & & & & \\ \bottomrule

\multicolumn{5}{l}{1 is the design by Lin et. al. \cite{Lin}}\\		
\multicolumn{5}{l}{2 is the design by Jayashree et. al.\cite{Jayashree}}	\\		
\end{tabular}
%}
\end{table}	

\begin{table}[tbhp]
\caption{T-count comparison of quantum \textit{Ctrl-Add} circuits}
\label{md-table:2}
\centering
\resizebox{\columnwidth}{!}{
\begin{tabular}{cccccccc}
%\multicolumn{8}{c}{}	\\ 
\midrule											
qubits 	&		& 	1	& 	2	&	Proposed	&		&	\% Impr.	&	\% 
Impr.	\\		
	&		& 		& 		&		&		&	w.r.t. 1	&	w.r.t. 2	
	\\	\cmidrule{1-1} \cmidrule{3-5} \cmidrule{7-8}	
4	&		&	224	&	119	&	98	&		&	56.25	&	17.65	\\		
8	&		&	448	&	231	&	182	&		&	59.38	&	21.21	\\		
16	&		&	896	&	455	&	350	&		&	60.94	&	23.08	\\		
32	&		&	1792	&	903	&	686	&		&	61.72	&	24.03	
\\		
64	&		&	3584	&	1799	&	1358	&		&	62.11	&	
24.51	\\		
128	&		&	7168	&	3591	&	2702	&		&	62.30	&	
24.76	\\		
256	&		&	14336	&	7175	&	5390	&		&	62.40	&	24.88	\\		
512	&		&	28672	&	14343	&	10766	&		&	62.45	&	24.94	\\	
1024	&		&	57344	&	28679	&	21518	&		&	62.48	&	24.97	\\
2048	&		&	114688	&	57351	&	43022	&		&	62.49	&	24.98	\\ \midrule
&		&	\multicolumn{3}{r}{Average:}					&		&	61.25	&	23.50	\\

\bottomrule	
									
\multicolumn{5}{l}{1 is the design by Lin et. al. \cite{Lin}}\\		
\multicolumn{5}{l}{2 is the design by Jayashree et. al.\cite{Jayashree}}	\\		

\end{tabular}
}
\end{table}

%\begin{table}[tbhp]
%\caption{T-depth comparison of quantum \textit{Ctrl-Add} circuits}
%\label{md-table:3}
%\centering
%\resizebox{\columnwidth}{!}{
%\begin{tabular}{cccccccc}
%%\multicolumn{8}{c}{T-depth comparison of quantum \textit{Ctrl-Add} circuits}	
%\midrule										
%qubits 	&		& 	1	& 	2	&	Proposed	&		&	\% Impr.	&	\% 
%Impr.	\\	
%	&		& 		& 		&		&		&	w.r.t. 1	&	w.r.t. 2	
%	\\	\cmidrule{1-1} \cmidrule{3-5} \cmidrule{7-8}
%4	&		&	16	&	31	&	14	&		&	12.50	&	54.84	\\	
%8	&		&	32	&	59	&	20	&		&	37.50	&	66.10	\\	
%16	&		&	64	&	115	&	36	&		&	43.75	&	68.70	\\	
%32	&		&	128	&	227	&	68	&		&	46.88	&	70.04	\\	
%64	&		&	256	&	451	&	132	&		&	48.44	&	70.73	\\	
%128	&		&	512	&	899	&	260	&		&	49.22	&	71.08	\\	
%%256	&		&	1024	&	1795	&	516	&		&	49.61	&	71.25	
%%\\	
%%512	&		&	2048	&	3587	&	1028	&		&	49.80	&	
%%71.34	\\	
%\bottomrule
%															
%\multicolumn{8}{l}{1 is the design in \cite{Lin}}\\		
%\multicolumn{8}{l}{2 is the design in \cite{Jayashree}}	\\		
%\end{tabular}
%}
%\end{table}

The T-count of the proposed quantum \textit{Ctrl-Add} circuit with no input carry is illustrated shortly for each step of the proposed design methodology.  We calculate total T-count for the proposed quantum \textit{Ctrl-Add} circuit with no input carry by summing the T-count for each step of the proposed design methodology.  

\begin {itemize} 

\item Step 1:
\begin{itemize}
\item The T-count for this Step is $0$.
%\item The circuit T-depth after this Step is $0$.
\end{itemize}

\item Step 2:
\begin{itemize}
\item The T-count for this step is $7$.
%\item The circuit T-depth after this step is 
%$3$.  This maximum T-depth is seen on the quantum register location that initially stores the value $a_{n-1}$. 
\end{itemize}

\item Step 3:
\begin{itemize}

\item The T-count for this Step is $7 \cdot (n-1)$.
%\item The circuit T-depth after this Step is $5$.  This maximum T-depth is seen on the quantum register locations that initially store the values $a_i$ where $1 \leq i \leq n-1$.
\end{itemize}

\item Step 4:
\begin{itemize}
\item The T-count for this Step is $28$.
%\item The circuit T-depth after this Step is $12$.  This maximum T-depth is seen on the quantum register location that initially stores the value $a_{n-1}$.

\end{itemize}

\item Step 5:
\begin{itemize}
\item The T-count for this Step is $14 \cdot (n-1)$.
%\item The circuit T-depth after this Step depends on the size of the proposed quantum \textit{Ctrl-Add} circuit with no input carry.  For $n \leq 5$, the circuit T-depth after this Step is $14$.  This maximum T-depth is seen on the quantum registered location that initially stores the value $a_n$.  For $n > 5$, the circuit T-depth after this Step is $2 \cdot n + 4$.  This maximum T-depth is seen on the quantum registered location that initially stores the value $ctrl$.  Thus, the T-depth of the proposed quantum \textit{Ctrl-Add} circuit with no input carry after this Step is defined as:
%
%% include formal equation here
%\begin{equation}
%\begin{cases}
%14 & \text{ if } n \leq 5 \\
%2 \cdot n + 4 & \text{ if } n > 5 \\
%\end{cases}
%\label{md-equation:10}
%\end{equation}
\end{itemize}

\item Step 6:
\begin{itemize}
\item The T-count for this Step is $0$.
%\item The circuit T-depth after this Step is the same as the circuit T-depth after Step 5 and is shown in equation \ref{md-equation:10}. 
\end{itemize}

\item Step 7:
\begin{itemize}
\item The T-count for this Step is $0$.
%\item The circuit T-depth after this Step is the same as the circuit T-depth after Step 5 and is shown in equation \ref{md-equation:10}.
\end{itemize}

\end {itemize}

Thus, the total T-count of an $n$ bit proposed quantum \textit{Ctrl-Add} circuit with no input carry is given as:

\begin{equation}
7 + 7 \cdot (n-1) + 28 + 14 \cdot (n-1) = 21 \cdot n + 14    
\label{md-equation:15}
\end{equation}

A comparison of the proposed quantum \textit{Ctrl-Add} circuit with no input carry with existing designs is illustrated in table \ref{md-table:1} which shows that the proposed design has a lower T-count compared to existing designs.  Table \ref{md-table:1} illustrates that the order of growth of the T-count for the proposed quantum \textit{Ctrl-Add} circuit with no input carry is linear.  Thus, the T-count is $\mathcal{O}(n)$.  Further, table \ref{md-table:1} shows that the savings in the T-count of the proposed design does not come at the cost of additional qubits.  Also, table \ref{md-table:1} illustrates that the order of growth of the qubit cost for the proposed quantum \textit{Ctrl-Add} circuit with no input carry is linear.  Thus, the qubit cost is $\mathcal{O}(n)$.

Table \ref{md-table:2} shows the comparison in terms of the T-count and shows that the proposed quantum \textit{Ctrl-Add} circuit with no input carry achieves improvement ratios ranging from $56.25 \%$ to $62.30 \%$ and $17.65 \%$ to $24.76 \%$ compared to the designs presented in Lin et. al. (\cite{Lin}) and Jayashree et. al. (\cite{Jayashree}) respectively.

%\section{Design Methodology of Proposed Quantum Adder-Subtractor Circuit with No Input Carry}
%\label{md_sub_add}

\section{Design of Proposed Quantum Integer Multiplication Circuit}
\label{md_mult}

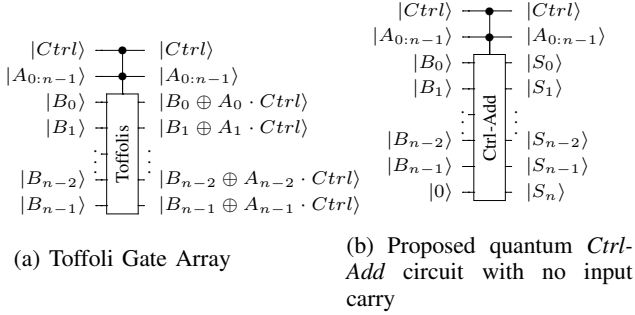
\begin{figure}[tbh]
\centering
\scriptsize
	\begin{subfigure}[th]{1.5in}

\[
\Qcircuit @C=0.3em @R=0.5em @!R{
\lstick{\ket{Ctrl}}	&\qw	&\ctrl{2}		&\qw		&\rstick{\ket{Ctrl}} \\
\lstick{\ket{A_{0:n-1}}}	&\qw	&\ctrl{1}				&\qw		&\rstick{\ket{A_{0:n-1}}} \\
\lstick{\ket{B_0}}	&\qw	&\multigate{4}{\begin{sideways}
Toffolis
\end{sideways}}				&\qw		&\rstick{\ket{B_0 \oplus A_0 \cdot Ctrl}} \\
\lstick{\ket{B_1}}	&\qw 	&\ghost{\begin{sideways}
Toffolis
\end{sideways}}				&\qw		&\rstick{\ket{B_1 \oplus A_1 \cdot Ctrl}} \\
\vdots & & \ghost{\begin{sideways}
Toffolis
\end{sideways}}	& & \vdots \\			
\lstick{\ket{B_{n-2}}}	&\qw	&\ghost{\begin{sideways}
Toffolis
\end{sideways}}		&\qw		& \rstick{\ket{B_{n-2} \oplus A_{n-2} \cdot Ctrl}} \\
\lstick{\ket{B_{n-1}}}	&\qw	&\ghost{\begin{sideways}
Toffolis \end{sideways}}			&\qw		& \rstick{\ket{B_{n-1} \oplus A_{n-1} \cdot Ctrl}} \\
}
\]
\caption{Toffoli Gate Array}	
\end{subfigure}	\qquad \qquad \begin{subfigure}[th]{1.5in}

\[
\Qcircuit @C=0.3em @R=0.5em @!R{
\lstick{\ket{Ctrl}}	&\qw	&\ctrl{2}		&\qw		&\rstick{\ket{Ctrl}} \\
\lstick{\ket{A_{0:n-1}}}	&\qw	&\ctrl{1}				&\qw		&\rstick{\ket{A_{0:n-1}}} \\
\lstick{\ket{B_0}} &\qw &\multigate{5}{\begin{sideways} Ctrl-Add \end{sideways}}			&\qw			&\rstick{\ket{S_0}} \\
\lstick{\ket{B_1}}	&\qw &\ghost{\begin{sideways} Ctrl-Add \end{sideways}}		&\qw		&\rstick{\ket{S_1}} \\
\vdots & & \ghost{\begin{sideways} Ctrl-Add \end{sideways}}	& & \vdots \\
\lstick{\ket{B_{n-2}}} &\qw &\ghost{\begin{sideways} Ctrl-Add \end{sideways}}			&\qw	&\rstick{\ket{S_{n-2}}} \\
\lstick{\ket{B_{n-1}}} &\qw &\ghost{\begin{sideways} Ctrl-Add \end{sideways}}			&\qw	&\rstick{\ket{S_{n-1}}} \\
\lstick{\ket{0}}  &\qw &\ghost{\begin{sideways} Ctrl-Add \end{sideways}}			&\qw	&\rstick{\ket{S_n}} \\
}
\]
  \caption{Proposed quantum \textit{Ctrl-Add} circuit with no input carry}
 \end{subfigure}
\caption{Graphical representation of components used in the proposed quantum integer multiplication circuit.}
%\caption{Quantum circuit implementations of the }
\label{mult:FIG4}
\end{figure}

The proposed quantum integer multiplication circuit is designed without garbage outputs and with a lower T-count compared to the existing design approaches which have no garbage outputs.  The quantum integer multiplication circuit is based on the proposed quantum \textit{Ctrl-Add} circuit with no input carry and the Toffoli gate array.  Figure \ref{mult:FIG4} shows the graphical representation of components used in the quantum integer multiplication circuit.

%include figure here someplace

 Consider the multiplication of $n$ two bit numbers $a$ and $b$ stored in quantum registers $\ket{A}$ and $\ket{B}$ respectively.  Further, consider a quantum register $\ket{P}$ where each qubit $\ket{P_i}$ where $0 \leq i \leq 2 \cdot n$ is initialized with $z = 0$.  At the end of the computation, the quantum registers $\ket{A}$ and $\ket{B}$ keep the values respectively.  Further, at the end of computation, the quantum register $\ket{P_i}$ that initially stored instances of the value $z$ will have the value $p_i$ for $0 \leq i \leq 2 \cdot n-1$.  Here, $p_i$ is the $i$th bit of the product of $a$ and $b$.  The quantum register location $\ket{P_{2 \cdot n}}$ is restored to the value $0$ at the end of computation.

The proposed design methodology reduces the T-count by using the proposed quantum \textit{Ctrl-Add} circuit with no input carry designed in section \ref{md_adder}.  Further, the proposed design methodology selectively replaces instances of the quantum \textit{Ctrl-Add} circuit with no input carry with a Toffoli gate array at the appropriate places. 

\begin{figure*}[t!bhp]
\centering
\small

	\begin{subfigure}[th]{1.35in}

\[
\Qcircuit @C=0.3em @R=0.5em @!R{
\lstick{\ket{B_0}}&	&\ctrl{5}			&\qw		&\rstick{\ket{B_0}} \\
\lstick{\ket{B_1}}&	&\qw			&\qw		&\rstick{\ket{B_1}} \\
\lstick{\ket{B_2}}&	&\qw			&\qw		&\rstick{\ket{B_2}} \\
\lstick{\ket{B_3}}&	&\qw		&\qw		&\rstick{\ket{B_3}} \\
\lstick{\ket{A_{3:0}}}&	&\ctrl{1}		&\qw		&\rstick{\ket{A_{3:0}}} \\
\lstick{\ket{0}}&	&\multigate{3}{\begin{sideways} Toffolis \end{sideways}}			&\qw		& \rstick{\ket{P_0}} \\
\lstick{\ket{0}}&	&\ghost{\begin{sideways} Toffolis 
\end{sideways}}			&\qw		&\rstick{\ket{B_0 \cdot A_1}} \\
\lstick{\ket{0}}&	&\ghost{\begin{sideways} Toffolis 
\end{sideways}}			&\qw		&\rstick{\ket{B_0 \cdot A_2}} \\
\lstick{\ket{0}}&	&\ghost{\begin{sideways} Toffolis 
\end{sideways}}			&\qw		&\rstick{\ket{B_0 \cdot A_3}} \\
\lstick{\ket{0}}&	&\qw			&\qw		&\rstick{\ket{0}} \\
\lstick{\ket{0}}&	&\qw			&\qw		&\rstick{\ket{0}} \\
\lstick{\ket{0}}&	&\qw			&\qw		&\rstick{\ket{0}} \\
\lstick{\ket{0}}&	&\qw		&\qw		& \rstick{\ket{0}} \\
\lstick{\ket{0}}&	&\qw		&\qw	& \rstick{\ket{0}}
}
\]	
\caption{After Step 1}		

	\end{subfigure} \qquad \begin{subfigure}[th]{1in}

\[
\Qcircuit @C=0.3em @R=0.5em @!R{
\lstick{\ket{B_0}}&	&\ctrl{5}			&\qw	&\qw &\qw	
&\rstick{\ket{B_0}} 
\\
\lstick{\ket{B_1}}&	&\qw			&\qw	&\ctrl{4} &\qw	
&\rstick{\ket{B_1}} 
\\
\lstick{\ket{B_2}}&	&\qw			&\qw	&\qw &\qw	&\rstick{\ket{B_2}} \\
\lstick{\ket{B_3}}&	&\qw		&\qw	&\qw &\qw	& \rstick{\ket{B_3}} \\
\lstick{\ket{A_{3:0}}}&	&\ctrl{1}		&\qw	&\ctrl{2} &\qw	
&\rstick{\ket{A_{3:0}}} \\
\lstick{\ket{0}}&	&\multigate{3}{\begin{sideways} Toffolis \end{sideways}}			&\qw	&\qw &\qw	& \rstick{\ket{P_0}} \\
\lstick{\ket{0}}&	&\ghost{\begin{sideways} Toffolis 
\end{sideways}}			&\qw	&\multigate{5}{ \begin{sideways} Ctrl-Add 
\end{sideways}} &\qw	&\rstick{\ket{P_1}} \\
\lstick{\ket{0}}&	&\ghost{\begin{sideways} Toffolis 
\end{sideways}}			&\qw	&\ghost{ \begin{sideways} Ctrl-Add 
\end{sideways}} &\qw	& \rstick{\ket{S_1}}\\
\lstick{\ket{0}}&	&\ghost{\begin{sideways} Toffolis 
\end{sideways}}			&\qw	&\ghost{ \begin{sideways} Ctrl-Add 
\end{sideways}} &\qw	&\rstick{\ket{S_2}} \\
\lstick{\ket{0}}&	&\qw			&\qw	&\ghost{ \begin{sideways} Ctrl-Add 
\end{sideways}} &\qw	&\rstick{\ket{S_3}} \\
\lstick{\ket{0}}&	&\qw			&\qw	&\ghost{ \begin{sideways} Ctrl-Add 
\end{sideways}} &\qw	& \rstick{\ket{S_4}} \\
\lstick{\ket{0}}&	&\qw			&\qw	&\ghost{ \begin{sideways} Ctrl-Add 
\end{sideways}} &\qw	& \rstick{\ket{0}} \\
\lstick{\ket{0}}&	&\qw		&\qw	&\qw &\qw	& \rstick{\ket{0}} \\
\lstick{\ket{0}}&	&\qw		&\qw	&\qw &\qw	& \rstick{\ket{0}}
}
\]

\caption{After Step 2}

\end{subfigure} \qquad \begin{subfigure}[th]{1.8in}
\small

\[
\Qcircuit @C=0.3em @R=0.5em @!R{
\lstick{\ket{B_0}}&	&\ctrl{5}			&\qw	&\qw &\qw &\qw &\qw	& 
\rstick{\ket{B_0}} \\
\lstick{\ket{B_1}}&	&\qw			&\qw	&\ctrl{4} &\qw	&\qw &\qw & 
\rstick{\ket{B_1}} \\
\lstick{\ket{B_2}}&	&\qw			&\qw	&\qw &\qw	&\ctrl{4} &\qw & 
\rstick{\ket{B_2}} \\
\lstick{\ket{B_3}}&	&\qw		&\qw	&\qw &\qw	&\qw &\qw & 
\rstick{\ket{B_3}} \\
\lstick{\ket{A_{3:0}}}&	&\ctrl{1}		&\qw	&\ctrl{2} &\qw	&\ctrl{3} &\qw 
& \rstick{\ket{A_{3:0}}}\\
\lstick{\ket{0}}&	&\multigate{3}{\begin{sideways} Toffolis \end{sideways}}			&\qw	&\qw &\qw	&\qw &\qw &\rstick{\ket{P_0}} \\
\lstick{\ket{0}}&	&\ghost{\begin{sideways} Toffolis 
\end{sideways}}			&\qw	&\multigate{5}{ \begin{sideways} Ctrl-Add 
\end{sideways}} &\qw	&\qw &\qw &\rstick{\ket{P_1}} \\
\lstick{\ket{0}}&	&\ghost{\begin{sideways} Toffolis 
\end{sideways}}			&\qw	&\ghost{ \begin{sideways} Ctrl-Add 
\end{sideways}} &\qw	&\multigate{5}{ \begin{sideways} Ctrl-Add 
\end{sideways}}  &\qw &\rstick{\ket{P_2}} \\
\lstick{\ket{0}}&	&\ghost{\begin{sideways} Toffolis 
\end{sideways}}			&\qw	&\ghost{ \begin{sideways} Ctrl-Add 
\end{sideways}} &\qw &\ghost{ \begin{sideways} Ctrl-Add \end{sideways}}  
&\qw 	& \rstick{\ket{S_1}} \\
\lstick{\ket{0}}&	&\qw			&\qw	&\ghost{ \begin{sideways} Ctrl-Add 
\end{sideways}} &\qw	&\ghost{ \begin{sideways} Ctrl-Add \end{sideways}} 
&\qw & 
\rstick{\ket{S_2}} \\
\lstick{\ket{0}}&	&\qw			&\qw	&\ghost{ \begin{sideways} Ctrl-Add 
\end{sideways}} &\qw	&\ghost{ \begin{sideways} Ctrl-Add \end{sideways}} 
&\qw & \rstick{\ket{S_3}} \\
\lstick{\ket{0}}&	&\qw			&\qw	&\ghost{ \begin{sideways} Ctrl-Add 
\end{sideways}} &\qw	&\ghost{ \begin{sideways} Ctrl-Add \end{sideways}} 
&\qw & \rstick{\ket{S_4}}\\
\lstick{\ket{0}}&	&\qw		&\qw	&\qw &\qw	&\ghost{ \begin{sideways} 
Ctrl-Add \end{sideways}} &\qw & \rstick{\ket{0}} \\
\lstick{\ket{0}}&	&\qw		&\qw	&\qw &\qw	&\qw &\qw & 
\rstick{\ket{0}}
}
\]

\caption{After First Iteration of Step 3}

\end{subfigure} \qquad \begin{subfigure}[th]{1.5in}

\small

\[
\Qcircuit @C=0.3em @R=0.5em @!R{
\lstick{\ket{B_0}}&	&\ctrl{5}			&\qw	&\qw &\qw &\qw &\qw &\qw &\qw	& \rstick{\ket{B_0}}\\
\lstick{\ket{B_1}}&	&\qw			&\qw	&\ctrl{4} &\qw	&\qw &\qw &\qw 
&\qw & \rstick{\ket{B_1}}\\
\lstick{\ket{B_2}}&	&\qw			&\qw	&\qw &\qw	&\ctrl{4} &\qw &\qw 
&\qw & \rstick{\ket{B_2}}\\
\lstick{\ket{B_3}}&	&\qw		&\qw	&\qw &\qw	&\qw &\qw &\ctrl{4} &\qw & 
\rstick{\ket{B_3}}\\
\lstick{\ket{A_{3:0}}}&	&\ctrl{1}		&\qw	&\ctrl{2} &\qw	&\ctrl{3} &\qw 
&\ctrl{4} &\qw & \rstick{\ket{A_{3:0}}} \\
\lstick{\ket{0}}&	&\multigate{3}{\begin{sideways} Toffolis \end{sideways}}			&\qw	&\qw &\qw	&\qw &\qw &\qw &\qw & \rstick{\ket{P_0}}\\
\lstick{\ket{0}}&	&\ghost{\begin{sideways} Toffolis 
\end{sideways}}			&\qw	&\multigate{5}{ \begin{sideways} Ctrl-Add 
\end{sideways}} &\qw	&\qw &\qw &\qw &\qw &\rstick{\ket{P_1}} \\
\lstick{\ket{0}}&	&\ghost{\begin{sideways} Toffolis 
\end{sideways}}			&\qw	&\ghost{ \begin{sideways} Ctrl-Add 
\end{sideways}} &\qw	&\multigate{5}{ \begin{sideways} Ctrl-Add 
\end{sideways}}  &\qw &\qw &\qw &\rstick{\ket{P_2}} \\
\lstick{\ket{0}}&	&\ghost{\begin{sideways} Toffolis 
\end{sideways}}			&\qw	&\ghost{ \begin{sideways} Ctrl-Add 
\end{sideways}} &\qw &\ghost{ \begin{sideways} Ctrl-Add \end{sideways}}  
&\qw 	&\multigate{5}{ \begin{sideways} Ctrl-Add \end{sideways}}  &\qw 
&\rstick{\ket{P_3}}\\
\lstick{\ket{0}}&	&\qw			&\qw	&\ghost{ \begin{sideways} Ctrl-Add \end{sideways}} &\qw	&\ghost{ \begin{sideways} Ctrl-Add \end{sideways}} &\qw &\ghost{\begin{sideways} Toffolis \end{sideways}} &\qw &\rstick{\ket{P_4}}\\
\lstick{\ket{0}}&	&\qw			&\qw	&\ghost{ \begin{sideways} Ctrl-Add \end{sideways}} &\qw	&\ghost{ \begin{sideways} Ctrl-Add \end{sideways}} &\qw &\ghost{\begin{sideways} Toffolis \end{sideways}} &\qw &\rstick{\ket{P_5}}\\
\lstick{\ket{0}}&	&\qw			&\qw	&\ghost{ \begin{sideways} Ctrl-Add 
\end{sideways}} &\qw	&\ghost{ \begin{sideways} Ctrl-Add \end{sideways}} 
&\qw &\ghost{\begin{sideways} Toffolis \end{sideways}} &\qw 
&\rstick{\ket{P_6}}\\
\lstick{\ket{0}}&	&\qw		&\qw	&\qw &\qw	&\ghost{ \begin{sideways} 
Ctrl-Add \end{sideways}} &\qw &\ghost{\begin{sideways} Ctrl-Add 
\end{sideways}} &\qw & \rstick{\ket{P_7}} \\
\lstick{\ket{0}}&	&\qw		&\qw	&\qw &\qw	&\qw &\qw &\ghost{ 
\begin{sideways} Ctrl-Add 
\end{sideways}} &\qw & 
\rstick{\ket{0}}
}
\]

\caption{After Second and Final Iteration of Step 3}

\end{subfigure}
\caption{Circuit Generation of quantum 4 bit integer multiplication circuit.  Steps 1-2 and all iterations of step 3.}
\label{mult:FIG5}
\end{figure*}
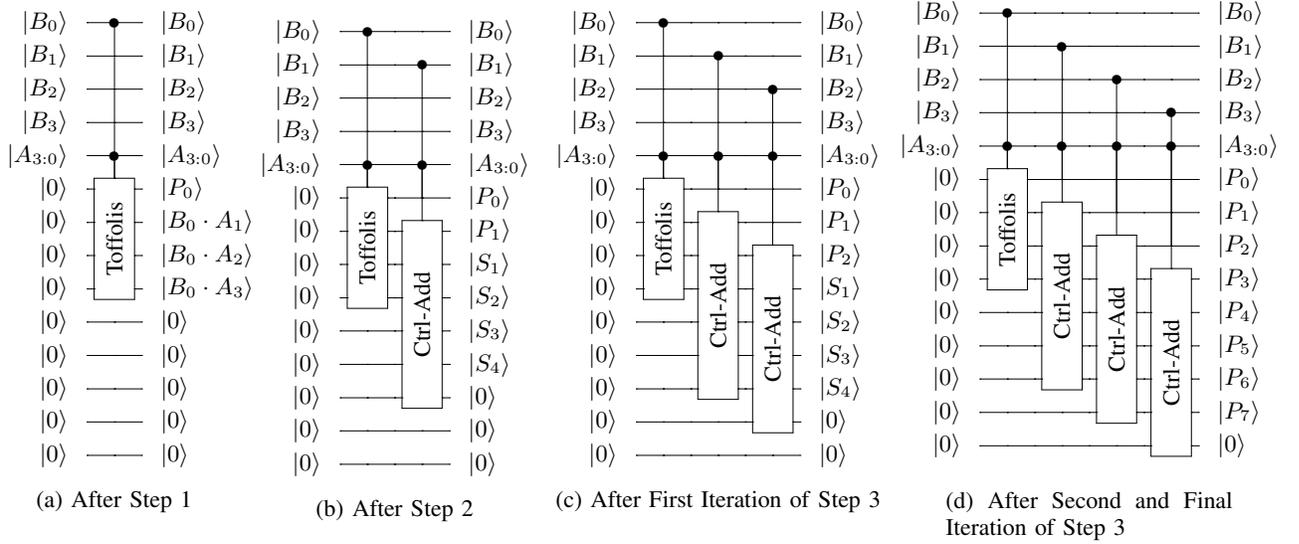

The proposed design methodology of generating the quantum integer multiplication circuit is explained below.  The proposed methodology is generic and can design a quantum integer multiplication circuit of any size.  The steps involved in the proposed methodology are presented for the conditional addition of two $n$ bit numbers $a_i$ and $b_i$, where $0 \leq i \leq n-1$.  An illustrative example of the generation of a quantum integer multiplication circuit that can perform the multiplication of two 4 bit numbers $a = a_0 \cdots a_3$ and $b = b_0 \cdots b_3$ is shown in figure \ref{mult:FIG5}.  The quantum circuit after each Step of the methodology is shown in figure \ref{mult:FIG5}.

\subsection{Steps of Design Methodology}

\begin {itemize}

\item Step 1: For $i = 0:1:n-1$

At locations $\ket{B_0}$,$\ket{A_i}$ and $\ket{P_i}$ apply a Toffoli gate such 
that the locations $\ket{B_0}$ and $\ket{A_i}$ will maintain the same value 
while location $\ket{P_i}$ transforms to the value $b_0 \cdot a_i$.  %The quantum integer multiplication circuit after Step 1 is shown for the multiplication of two 4-bit numbers in figure \ref{mult:FIG5}.

\item Step 2: Step 2 has the following three sub-steps

\begin {itemize}

\item Step 1: For $i = 1:1:n$

Apply the pair of locations $\ket{A_{i-1}}$ and $\ket{P_i}$ to a quantum 
\textit{Ctrl-Add} circuit with no input carry such that the location 
$\ket{A_{i-1}}$ will maintain the same value while location $\ket{P_i}$ 
transforms to the sum bit $s_{i-1}$.

\item Step 2: Apply location $\ket{B_1}$ to the quantum \textit{Ctrl-Add} 
circuit such that the operation of the \textit{Ctrl-Add} circuit will be 
conditioned on the value of $\ket{B_1}$.

\item Step 3: Apply locations $\ket{P_{n+1}}$ and $\ket{P_{n+2}}$ to the 
quantum \textit{Ctrl-Add} circuit such that location $\ket{P_{n+1}}$ 
transforms to the sum bit $s_n$ and location $\ket{P_{n+2}}$ will maintain the 
same value at the end of computation.  

\end {itemize}

% The quantum integer multiplication circuit after Step 2 is shown for the multiplication of two 4-bit numbers in figure \ref{mult:FIG5}.

\item Step 3:  Step 3 is repeated $n-2$ times. For $j = 2:1:n-1$:  This Step has the following three sub-steps

\begin {itemize}

\item Step 1: For $i = 0:1:n-1$

Apply the pair of locations $\ket{A_i}$ and $\ket{P_{i+j}}$ to a quantum 
\textit{Ctrl-Add} circuit with no input carry such that the location 
$\ket{A_i}$ will maintain the same value while location $\ket{P_{i+j}}$ 
transforms to the sum bit $s_i$.

\item Step 2: Apply location $\ket{B_j}$ to the quantum \textit{Ctrl-Add} 
circuit such that the operation of the \textit{Ctrl-Add} circuit will be 
conditioned on the value of $\ket{B_j}$.

\item Step 3: Apply locations $\ket{P_{n+j}}$ and $\ket{P_{n+j+1}}$ to the 
quantum \textit{Ctrl-Add} circuit such that location $\ket{P_{n+j}}$ 
transforms to the sum bit $s_n$ and location $\ket{P_{n+j+1}}$ will maintain 
the same value at the end of computation.  

\end {itemize}

% The quantum integer multiplication circuit after each iteration of Step 3 is shown for the multiplication of two 4-bit numbers in figure \ref{mult:FIG5}.

\end{itemize}

\textbf{Theorem:} \textit{Let $a$ and $b$ be two $n$ bit binary numbers represented as $a_i$ and $b_i$ and $p$ is another $2 \cdot n$ bit binary number represented as $p_i$ and $p_i = 0$, then the proposed design methodology results in a quantum integer multiplication circuit that functions correctly.  The proposed design methodology designs an $n$ bit multiplication circuit that that produces the product output at the quantum register where $p_i$ is initially stored while the locations where $a$ and $b$ are initially stored are restored to the values $a$ and $b$ respectively.}

\textbf{Proof:} The proposed design methodology will make the following changes on the inputs.  For illustrative purposes, the transformation of the input states for a four bit quantum integer multiplication circuit after each Step is shown in figure \ref{mult:FIG5}.:

\begin{itemize}

\item Step 1: Step 1 of the proposed methodology transforms the input states to:

\end{itemize}

\begin{equation}
\left( \bigotimes_{i = 0}^{n-1} \ket{A_i} \right) \left( \bigotimes_{i = 0}^{n-1} \ket{B_i} \right) \left( \bigotimes_{i = 0}^{n-1} \ket{B_0 \cdot A_i} \right) \left( \bigotimes_{i = n}^{2 \cdot n} \ket{0} \right)
\label{md-equation:20}
\end{equation}

\begin{itemize}

%\item[] For illustrative purposes, the transformation of the input states for a four bit quantum integer multiplication circuit after Step 1 is shown in figure \ref{mult:FIG5}.

\item Step 2: Step 2 has a \textit{Ctrl-Add} circuit with no input carry which takes the inputs $a_i$ and $p_{i+1}$ for $0 \leq i \leq n-1$ and the input $b_1$.  At the end of computation the locations $p_{i+1}$ will have the value $s_i$ for $0 \leq i \leq n-1$.  Further, at the end of computation the locations which originally stored $b_1$ and $a_i$ are restored to the values $b_1$ and $a_i$, respectively where $0 \leq i \leq n-1$.  Thus, after Step 2 the input states are transformed to:

\end{itemize}

\begin{equation}
\left( \bigotimes_{i = 0}^{n-1} \ket{A_i} \right) \left( \bigotimes_{i = 0}^{n-1} \ket{B_i} \right) \ket{p_0} \left( \bigotimes_{i = 1}^{n+1} \ket{s_{i-1}} \right) \left( \bigotimes_{i = {n+2}}^{2 \cdot n} \ket{0} \right)
\label{md-equation:21}
\end{equation}

\begin{itemize}

%\item[] For illustrative purposes, the transformation of the inputs states of a four bit quantum integer multiplication circuit after Step 2 is shown in figure \ref{mult:FIG5}.

\item Step 3: Step 3 is repeated a total of $n-2$ times.  

\begin{itemize}

\item For Iteration one of Step 3, a quantum \textit{Ctrl-Add} circuit with no input carry takes the inputs as $a_i$ and $p_{i+2}$ for $0 \leq i \leq n-1$ and the input $b_2$.  At the end of computation the locations $p_{i+2}$ will have the value $s_i$ for $0 \leq i \leq n-1$.  Further, at the end of computation the locations which originally stored $b_2$ and $a_i$ are restored to the values $b_2$ and $a_i$, respectively where $0 \leq i \leq n-1$.  Thus, after Iteration one of Step 3 the input states are transformed to:

\end{itemize}
\end{itemize}

\begin{equation}
\left( \bigotimes_{i = 0}^{n-1} \ket{A_i} \right) \left( \bigotimes_{i = 0}^{n-1} \ket{B_i} \right) \ket{p_0} \ket{p_1} \left( \bigotimes_{i = 2}^{n+2} \ket{s_{i-2}} \right) \left( \bigotimes_{i = {n+3}}^{2 \cdot n} \ket{0} \right)
\label{md-equation:22}
\end{equation}

\begin{itemize}
\item[]
\begin{itemize}

%\item[] An example of the transformation of the input states after the first iteration of step 3 is illustrated for a 4-bit quantum integer multiplication circuit in figure \ref{mult:FIG5}.

\item For Iteration $j$ of Step 3 where $2 \leq j \leq n-3$, a quantum \textit{Ctrl-Add} circuit with no input carry takes the inputs as $a_i$ and $p_{i+1+j}$ for $0 \leq i \leq n-1$ and the input $b_{1+j}$. At the end of computation the locations $p_{i+1+j}$ will have the value $s_i$ for $0 \leq i \leq n-1$.  Further, at the end of computation the locations which originally stored $b_{1+j}$ and $a_{i}$ are restored to the values $b_{1+j}$ and $a_{i}$, respectively, where $0 \leq i \leq n-1$.  Thus, after Iteration $j$ of Step 3 the input states $\ket{A_i}$ and $\ket{B_i}$ for $0 \leq i \leq n-1$ are transformed to:

\end{itemize}
\end{itemize}

\begin{equation}
\left( \bigotimes_{i = 0}^{n-1} \ket{A_i} \right) \left( \bigotimes_{i = 0}^{n-1} \ket{B_i} \right)
\label{md-equation:23-a}
\end{equation}

\begin{itemize}
\item[]
\begin{itemize}
\item[] After Iteration $j$ of Step 3 the input states $\ket{P_i}$ for $0 \leq i \leq 2 \cdot n$ is transformed to:
\end{itemize}
\end{itemize}

\begin{equation}
 \left( \bigotimes_{i = 0}^{j} \ket{P_i} \right)  \left( \bigotimes_{i = j+1}^{j+1+n} \ket{s_{i-j-1}} \right) \left( \bigotimes_{i = {j+n+2}}^{2 \cdot n} \ket{0} \right)
\label{md-equation:23-b}
\end{equation}

\begin{itemize}
\item[]
\begin{itemize}

\item For Iteration $n-2$ of Step 3, a quantum \textit{Ctrl-Add} circuit with no input carry takes the inputs as $a_i$ and $p_{i+(n-2)}$ for $0 \leq i \leq n-1$ and the input $b_{n-1}$.  At the end of computation the locations $p_{i+(n-2)}$ will have the value $s_i$ for $0 \leq i \leq n-1$.  Further, at the end of computation the locations which originally stored $b_{n-1}$ and $a_{i}$ are restored to the values $b_{n-1}$ and $a_{i}$, respectively, where $0 \leq i \leq n-1$. 

\begin{equation}
\left( \bigotimes_{i = 0}^{n-1} \ket{A_i} \right) \left( \bigotimes_{i = 0}^{n-1} \ket{B_i} \right) \left( \bigotimes_{i = 0}^{2 \cdot n -1} \ket{P_i} \right)  \ket{0} 
\label{md-equation:24}
\end{equation}

%An example of the transformation of the input states after the $n-2$ iteration of step 3 is illustrated for a 4-bit quantum integer multiplication circuit in figure \ref{mult:FIG5}.

\end{itemize}

\end{itemize}

Thus, the proposed design methodology transforms the quantum register $\ket{P_i}$ that originally stored $0$ to the product of $a$ and $b$ for $0 \leq i \leq 2 \cdot n -1$ while the quantum registers $\ket{A}$ and $\ket{B}$ where $a$ and $b$ are initially stored, respectively, will be restored to the values $a$ and $b$, respectively.  Further, the quantum register location $\ket{P_{2 \cdot n}}$ where $0$ was stored initially is restored to the value $0$.  This proves to the correctness of the proposed methodology to design a quantum integer multiplication circuit.

\subsection{T-count Analysis}

\begin{table*}[bthp]
\centering
\caption{Comparison of quantum integer multiplication circuits}
\label{md-table:4}
%\resizebox{\textwidth}{!}{
\begin{tabular}{cccccc}														
	\midrule										
	&		& 		1		& 		2		& 		3		&		Proposed		\\ \cmidrule{1-1} \cmidrule{3-6}
T-count		& &		$	56 \cdot n^2	$	&	$	28 \cdot n^2 + 7 \cdot n	$	&	$	42 \cdot n^2 - 42 \cdot n +48	$	&	$	21 \cdot n^2 - 14	$	\\
%T depth		& &		$	14 \cdot n	$	&	$	24 \cdot n	$	&	$	8 \cdot n	$	&	$	14 \cdot n - 11	$ \\	
qubits 	&		&	$5 \cdot n + 1$			&		$4 \cdot n + 1$		&		NA		&		$4 \cdot n + 1$		\\
%garbage &  & 0  & 0 & NA &  0 \\
ancillae &  & $3 \cdot n + 1$  & $2 \cdot n + 1$ & NA &  $2 \cdot n + 1$ \\

\bottomrule															

\multicolumn{6}{l}{1 is the design by Lin et. al. \cite{Lin}}	\\
\multicolumn{6}{l}{2 is the design by Jayashree et. al. \cite{Jayashree}}	\\
\multicolumn{6}{l}{3 is the design by Babu \cite{Babu} modified to remove garbage output.} \\
\multicolumn{6}{l}{Table entries are marked NA where a closed-form expression is not available for the design by Babu \cite{Babu}.}		\\
											
\\																	

\end{tabular}
%}	
\end{table*}

\begin{table*}[bthp]
\centering
\caption{T-count comparison of quantum integer multiplication circuits}
\label{md-table:5}
%\resizebox{\textwidth}{!}{
\begin{tabular}{cccccccccc}
	\midrule																	
qubits 	&		& 	1	& 	2	&	3	&			Proposed	&		&	\% Impr.	&	\% Impr.	&	\% Impr.			\\	
	&		& 		& 		&		&				&		&	w.r.t. 1	&	w.r.t. 2	&	w.r.t. 3			\\	\cmidrule{1-1} \cmidrule{3-6} \cmidrule{8-10}
4	&		&	896	&	476	&	528	&			322	&		&	64.06	&	32.35	&	39.02			\\	
8	&		&	3584	&	1848	&	2352	&			1330	&		&	62.89	&	28.03	&	43.45			\\	
16	&		&	14336	&	7280	&	10032	&			5362	&		&	62.60	&	26.35	&	46.55			\\	
32	&		&	57344	&	28896	&	41520	&			21490	&		&	62.52	&	25.63	&	48.24			\\	
64	&		&	229376	&	115136	&	169008	&			86002	&		&	62.51	&	25.30	&	49.11			\\	
128	&		&	917504	&	459648	&	682032	&			344050	&		&	62.50	&	25.15	&	49.56			\\	
256	&		&	3670016	&	1836800	&	2740272	&			1376242	&		&	62.50	&	25.07	&	49.78			\\	
512	&		&	14680064	&	7343616	&	10985520	&			5505010	&		&	62.50	&	25.04	&	49.89			\\	
1024	&		&	58720256	&	29367296	&	43991088	&			22020082	&		&	62.50	&	25.02	&	49.94		\\	
2048	&		&	234881024	&	117454848	&	176062512	&			88080370	&		&	62.50	&	25.01	&	49.97		\\	\midrule
  &		&	\multicolumn{4}{r}{Average:}									&		&	62.71	&	26.30	&	47.55			\\	\bottomrule

\multicolumn{10}{l}{1 is the design by Lin et. al. \cite{Lin}}	\\
\multicolumn{10}{l}{2 is the design by Jayashree et. al. \cite{Jayashree}}	\\
\multicolumn{10}{l}{3 is the design by Babu \cite{Babu} modified to remove garbage output.} \\											
\end{tabular}	
%}
\end{table*}

The T-count of the proposed quantum integer multiplication circuit is illustrated shortly for each step of the proposed design methodology.  We calculate total T-count for the proposed quantum integer multiplication circuit by summing the T-count for each step of the proposed design methodology.  

\begin{itemize}

\item[]
%\item[]
%\item[]

\item Step 1:
\begin{itemize}
\item The T-count for this Step is $7 \cdot n$.
\end{itemize}

\item Step 2:
\begin{itemize}
\item The T-count for this Step is $28 \cdot n + 14$.

\end{itemize}

% include formal equation here

\item Step 3: Step 3 is repeated $n-2$ times.  For $i = 0:1:n-2$:
\begin {itemize}
\item The T-count for the $i$th iteration of this Step is $28 \cdot n + 14$.
\end{itemize}

% include formal equation here

\end {itemize}

Therefore, the total T-count of the proposed quantum integer multiplication circuit is $21 \cdot n^2 -14$.  

A comparison of the proposed quantum integer multiplication circuit with existing designs is illustrated in table \ref{md-table:4} which shows that the proposed design has a lower T-count compared to existing designs.  To compare the proposed work against the recent quantum integer multiplication circuit design presented by Babu \cite{Babu}, we implemented the design using the fault tolerant Clifford + T gate family.  We used the Clifford + T implementation of the square root of not gate presented in \cite{Maslov} in the implementation of the design by Babu.  The implementation in \cite{Maslov} requires three T gates and four Clifford gates.  We also made the quantum integer multiplication circuit designed by Babu garbageless by applying the Bennett's garbage removal scheme described in section \ref{md_ref_past_stuff} .  Consequently, the garbageless form of the quantum circuit by Babu requires $2 \cdot n + 1$ additional qubits and sees an increase in the T-count by a factor of 2x. 

Table \ref{md-table:4} illustrates that the order of growth of the T-count for the proposed quantum integer multiplication circuit is quadratic.  Thus, the T-count is $\mathcal{O}(n^2)$.  Table \ref{md-table:4} shows that the proposed design has a low overall qubit cost.  Table \ref{md-table:4} illustrates that the order of growth of the qubits for the proposed quantum integer multiplication circuit is linear.  Thus, the qubit cost is $\mathcal{O}(n)$.

\begin{table}[tbhp]
\centering
\caption{Comparison between the design by Babu \cite{Babu} and the proposed work in terms of ancillae  }
\label{md-table:6}
%\resizebox{\columnwidth}{!}{
\begin{tabular}{cccccc}
	\midrule																	
						
qubits	& &	1	&	 Proposed &	&	\% Impr.	\\	
	& & &	&		&	w.r.t. 1	\\	\cmidrule{1-1} \cmidrule{3-4} \cmidrule{6-6}  
4	& &	18	&	9	& &	50.00	\\
8	& &	57	&	17	& &	70.18	\\
16	& &	178	&	33	& &	81.46	\\
32	& &	608	&	65	& &	89.31	\\
64	& &	2210	& 	129	& &	94.16	\\
128	& &	8368	& 	257	& &	96.93	\\  \midrule
 &		&	\multicolumn{3}{r}{Average:}		&		80.34	\\ \bottomrule
											
\multicolumn{6}{l}{1 is the design by Babu \cite{Babu} modified to} \\
\multicolumn{6}{l}{remove garbage output} \\	
\multicolumn{6}{l}{The designs by Lin et. al. and Jayashree et. al. are not } \\
\multicolumn{6}{l}{compared because the ancillae for these designs is} \\			
\multicolumn{6}{l}{nearly the same as the proposed work.} \\

\end{tabular}	
%}
\end{table}

%\end{table}																	

%\begin{table}[tbhp]
%\centering
%\caption{T-depth comparison of quantum integer multiplication circuits}
%\label{md-table:6}
%\resizebox{\columnwidth}{!}{
%\begin{tabular}{cccccccccc}
% \midrule										
%qubits 	&		& 	1	& 	2	&	3	&	Proposed	&		&	\% Impr.	&	\% Impr.	&	\% Impr.	\\					
%	&		& 		& 		&		&		&		&	w.r.t. 1	&	w.r.t. 2	&	w.r.t. 3	\\	\cmidrule{1-1} \cmidrule{3-6} \cmidrule{8-10}				
%4	&		&	56	&	96	&	32	&	45	&		&	19.64	&	53.13	&	-28.89	\\
%8	&		&	112	&	192	&	64	&	101	&		&	9.82	&	47.40	&	-36.63	\\
%16	&		&	224	&	384	&	128	&	213	&		&	4.91	&	44.53	&	-39.91	\\
%32	&		&	448	&	768	&	256	&	437	&		&	2.46	&	43.10	&	-41.42	\\
%64	&		&	896	&	1536	&	512	&	885	&		&	1.23	&	42.38	&	-42.15	\\
%128	&		&	1792	&	3072	&	1024	&	1781	&		&	0.61	&	42.02	&	-42.50	\\
%%256	&		&	3584	&	6144	&	2048	&	3573	&		&	0.31	&	41.85	&	-42.68	\\
%%512	&		&	7168	&	12288	&	4096	&	7157	&		&	0.15	&	41.76	&	-42.77	\\
%	
%		\bottomrule				
%																								
%\multicolumn{10}{l}{1 is the design in \cite{Lin}}			\\																			
%\multicolumn{10}{l}{2 is the design in \cite{Jayashree}} \\																					
%\multicolumn{10}{l}{3 is the design in \cite{Babu} modified to remove garbage output.} \\
%\multicolumn{10}{l}{ Design 3 requires at least $47$ \% more ancillae than proposed work.}																							\\	
%	
%\end{tabular}	
%}
%\end{table}	

Table \ref{md-table:5} shows the comparison in terms of T-count which shows that the proposed design methodology achieves improvement ratios ranging from $39.02 \%$ to $49.56 \%$, $62.50 \%$ to $64.06 \%$ and $25.15 \%$ to $32.35 \%$ compared to the designs presented by Babu (\cite{Babu}), Lin et. al. (\cite{Lin}) and Jayashree et. al.(\cite{Jayashree}).  Table \ref{md-table:6} shows the comparison in terms of ancillae which shows that the proposed design methodology achieves improvement ratios ranging from $50.00 \%$ to $96.93 \%$ compared to the recently proposed design by Babu (\cite{Babu}).  Table \ref{md-table:7} shows the comparison in terms of total qubits and shows that the proposed design methodology achieves improvement ratios ranging from $59.52 \%$ to $94.22 \%$ compared to the recently proposed design by Babu (\cite{Babu}).  We calculated total qubits by summing the garbage outputs, qubits holding the product and qubits of the inputs.

\begin{table}[tbhp]
\centering
\caption{Comparison between the design by Babu \cite{Babu} and the proposed work in terms of total qubits  }
\label{md-table:7}
%\resizebox{\columnwidth}{!}{
	\begin{tabular}{cccccc}
		\midrule																	
		
		qubits	& &	1	&	Proposed &	&	\% Impr.	\\	
		 & & & &			&	w.r.t. 1	 \\	\cmidrule{1-1} \cmidrule{3-4} \cmidrule{6-6}  
		4	& &	42	&	17	& &	59.52	\\
		8	& &	90	&	33	& &	63.33	\\
		16	& &	243	&	65	& &	73.25	\\
		32	& &	737	&	129	& &	82.50	\\
		64	& &	2467	&	257	& &	89.58	\\
		128	& &	8881	&	513	& &	94.22	\\ \midrule
 &		&	\multicolumn{3}{r}{Average:}		&		77.07
	\\ \bottomrule

\multicolumn{6}{l}{1 is the design by Babu \cite{Babu} modified to} \\
\multicolumn{6}{l}{remove garbage output} \\		
\multicolumn{6}{l}{total qubits is the sum of the garbage outputs, qubits} \\ 
\multicolumn{6}{l}{for the product and the qubits of the inputs } \\
\multicolumn{6}{l}{The designs by Lin et. al. and Jayashree et. al. are not } \\
\multicolumn{6}{l}{compared because the total qubits for these designs is} \\			
\multicolumn{6}{l}{nearly the same as the proposed work.} \\

	\end{tabular}	
%}
\end{table}

\section{Conclusions}
\label{md_done}

In this work, we have presented the design of a quantum integer multiplication circuit optimized in terms of T-count with $4 \cdot n + 1$ qubits.  Further, the proposed quantum integer multiplication circuit does not produce garbage outputs.  The design of subcomponents used in the proposed quantum integer multiplication circuit such as the proposed quantum conditional addition circuit with no input carry are also illustrated.  The proposed quantum integer multiplication circuit is shown to be superior to the existing designs in terms of T-count.  All of the proposed designs in this work are functionally verified by formal proof and Verilog simulation.  We conclude that the proposed quantum integer multiplication circuit will find applications in quantum computing that requires integer multiplication where T-count is of primary concern.

\bibliographystyle{IEEEtran}
% argument is your BibTeX string definitions and bibliography database(s)
\bibliography{MULTbiblio.bib}

% Generated by IEEEtran.bst, version: 1.13 (2008/09/30)
\begin{thebibliography}{10}
\providecommand{\url}[1]{#1}
\csname url@samestyle\endcsname
\providecommand{\newblock}{\relax}
\providecommand{\bibinfo}[2]{#2}
\providecommand{\BIBentrySTDinterwordspacing}{\spaceskip=0pt\relax}
\providecommand{\BIBentryALTinterwordstretchfactor}{4}
\providecommand{\BIBentryALTinterwordspacing}{\spaceskip=\fontdimen2\font plus
\BIBentryALTinterwordstretchfactor\fontdimen3\font minus
  \fontdimen4\font\relax}
\providecommand{\BIBforeignlanguage}[2]{{%
\expandafter\ifx\csname l@#1\endcsname\relax
\typeout{** WARNING: IEEEtran.bst: No hyphenation pattern has been}%
\typeout{** loaded for the language `#1'. Using the pattern for}%
\typeout{** the default language instead.}%
\else
\language=\csname l@#1\endcsname
\fi
#2}}
\providecommand{\BIBdecl}{\relax}
\BIBdecl

\bibitem{Babu}
\BIBentryALTinterwordspacing
H.~M.~H. Babu, ``Cost-efficient design of a quantum multiplier--accumulator
  unit,'' \emph{Quantum Information Processing}, vol.~16, no.~1, p.~30, 2016.
  [Online]. Available: \url{http://dx.doi.org/10.1007/s11128-016-1455-0}
\BIBentrySTDinterwordspacing

\bibitem{Lin}
\BIBentryALTinterwordspacing
C.-C. Lin, A.~Chakrabarti, and N.~K. Jha, ``Qlib: Quantum module library,''
  \emph{J. Emerg. Technol. Comput. Syst.}, vol.~11, no.~1, pp. 7:1--7:20, Oct.
  2014. [Online]. Available: \url{http://doi.acm.org/10.1145/2629430}
\BIBentrySTDinterwordspacing

\bibitem{Jayashree}
\BIBentryALTinterwordspacing
H.~V. Jayashree, H.~Thapliyal, H.~R. Arabnia, and V.~K. Agrawal,
  ``Ancilla-input and garbage-output optimized design of a reversible quantum
  integer multiplier,'' \emph{The Journal of Supercomputing}, vol.~72, no.~4,
  pp. 1477--1493, 2016. [Online]. Available:
  \url{http://dx.doi.org/10.1007/s11227-016-1676-0}
\BIBentrySTDinterwordspacing

\bibitem{Cheung}
D.~Cheung, D.~Maslov, J.~Mathew, and D.~K. Pradhan, \emph{Theory of Quantum
  Computation, Communication, and Cryptography: Third Workshop, TQC 2008 Tokyo,
  Japan, January 30 - February 1, 2008. Revised Selected Papers}.\hskip 1em
  plus 0.5em minus 0.4em\relax Berlin, Heidelberg: Springer Berlin Heidelberg,
  2008, ch. On the Design and Optimization of a Quantum Polynomial-Time Attack
  on Elliptic Curve Cryptography, pp. 96--104.

\bibitem{Bowregard}
S.~Beauregard, ``{Circuit for Shor's algorithm using 2n+3 gubits},''
  \emph{{QUANTUM INFORMATION \& COMPUTATION}}, vol.~{3}, no.~{2}, pp.
  {175--185}, {MAR} {2003}.

\bibitem{Montanaro}
A.~{Montanaro}, ``{Quantum pattern matching fast on average},'' \emph{ArXiv
  e-prints}, Aug. 2014.

\bibitem{Shparlinski}
W.~van Dam and I.~E. Shparlinski, \emph{Classical and Quantum Algorithms for
  Exponential Congruences}.\hskip 1em plus 0.5em minus 0.4em\relax Berlin,
  Heidelberg: Springer Berlin Heidelberg, 2008, pp. 1--10.

\bibitem{Proos}
J.~Proos and C.~Zalka, ``\BIBforeignlanguage{{English}}{{Shor's discrete
  logarithm quantum algorithm for elliptic curves}},''
  \emph{\BIBforeignlanguage{{English}}{{QUANTUM INFORMATION \& COMPUTATION}}},
  vol.~{3}, no.~{4}, pp. {317--344}, {JUL} {2003}.

\bibitem{Seroussi}
W.~{van Dam} and G.~{Seroussi}, ``{Efficient Quantum Algorithms for Estimating
  Gauss Sums},'' \emph{eprint arXiv:quant-ph/0207131}, Jul. 2002.

\bibitem{Quipper}
P.~S. et. al., \emph{The Quipper System}, 2016, available at:
  http://www.mathstat.dal.ca/~selinger/quipper/doc/.

\bibitem{LIQUi}
D.~W. et. al., \emph{Language-Integrated Quantum Operations: LIQUi|>}, 2016,
  available at:
  https://www.microsoft.com/en-us/research/project/language-integrated-quantum-operations-liqui/.

\bibitem{Maslov34}
D.~{Maslov}, ``{Basic circuit compilation techniques for an ion-trap quantum
  machine},'' \emph{ArXiv e-prints}, Mar. 2016.

\bibitem{Haner}
T.~{H{\"a}ner}, D.~S. {Steiger}, K.~{Svore}, and M.~{Troyer}, ``{A Software
  Methodology for Compiling Quantum Programs},'' \emph{ArXiv e-prints}, Apr.
  2016.

\bibitem{Fredkin}
\BIBentryALTinterwordspacing
E.~Fredkin and T.~Toffoli, ``Conservative logic,'' \emph{International Journal
  of Theoretical Physics}, vol.~21, no.~3, pp. 219--253, 1982. [Online].
  Available: \url{http://dx.doi.org/10.1007/BF01857727}
\BIBentrySTDinterwordspacing

\bibitem{Webster}
\BIBentryALTinterwordspacing
P.~Webster, S.~D. Bartlett, and D.~Poulin, ``Reducing the overhead for quantum
  computation when noise is biased,'' \emph{Phys. Rev. A}, vol.~92, p. 062309,
  Dec 2015. [Online]. Available:
  \url{http://link.aps.org/doi/10.1103/PhysRevA.92.062309}
\BIBentrySTDinterwordspacing

\bibitem{Zhou}
\BIBentryALTinterwordspacing
X.~Zhou, D.~W. Leung, and I.~L. Chuang, ``Methodology for quantum logic gate
  construction,'' \emph{Phys. Rev. A}, vol.~62, p. 052316, Oct 2000. [Online].
  Available: \url{http://link.aps.org/doi/10.1103/PhysRevA.62.052316}
\BIBentrySTDinterwordspacing

\bibitem{Paler_DAC}
A.~Paler and S.~J. Devitt, ``An introduction into fault-tolerant quantum
  computing,'' in \emph{2015 52nd ACM/EDAC/IEEE Design Automation Conference
  (DAC)}, June 2015, pp. 1--6.

\bibitem{Polian_DAC}
I.~Polian and A.~G. Fowler, ``Design automation challenges for scalable quantum
  architectures,'' in \emph{2015 52nd ACM/EDAC/IEEE Design Automation
  Conference (DAC)}, June 2015, pp. 1--6.

\bibitem{Mosca2}
M.~Amy, D.~Maslov, and M.~Mosca, ``Polynomial-time t-depth optimization of
  clifford+t circuits via matroid partitioning,'' \emph{IEEE Transactions on
  Computer-Aided Design of Integrated Circuits and Systems}, vol.~33, no.~10,
  pp. 1476--1489, Oct 2014.

\bibitem{Miller}
D.~Miller, M.~Soeken, and R.~Drechsler, ``Mapping ncv circuits to optimized
  clifford+t circuits,'' in \emph{Reversible Computation}, ser. Lecture Notes
  in Computer Science, S.~Yamashita and S.-i. Minato, Eds.\hskip 1em plus 0.5em
  minus 0.4em\relax Springer International Publishing, 2014, vol. 8507, pp.
  163--175.

\bibitem{Maslov}
M.~Amy, D.~Maslov, M.~Mosca, and M.~Roetteler, ``A meet-in-the-middle algorithm
  for fast synthesis of depth-optimal quantum circuits,'' \emph{IEEE
  Transactions on Computer-Aided Design of Integrated Circuits and Systems},
  vol.~32, no.~6, pp. 818--830, June 2013.

\bibitem{Gosset}
D.~Gosset, V.~Kliuchnikov, M.~Mosca, and V.~Russo, ``An algorithm for the
  t-count,'' \emph{ArXiv e-prints}, Aug. 2013.

\bibitem{Lidia}
L.~{Ruiz-Perez} and J.~C. {Garcia-Escartin}, ``{Quantum arithmetic with the
  Quantum Fourier Transform},'' \emph{ArXiv e-prints}, Nov. 2014.

\bibitem{Haghparast}
\BIBentryALTinterwordspacing
M.~HAGHPARAST, M.~MOHAMMADI, K.~NAVI, and M.~ESHGHI, ``Optimized reversible
  multiplier circuit,'' \emph{Journal of Circuits, Systems and Computers},
  vol.~18, no.~02, pp. 311--323, 2009. [Online]. Available:
  \url{http://www.worldscientific.com/doi/abs/10.1142/S0218126609005083}
\BIBentrySTDinterwordspacing

\bibitem{Akbar}
\BIBentryALTinterwordspacing
E.~P.~A. Akbar, M.~Haghparast, and K.~Navi, ``Novel design of a fast reversible
  wallace sign multiplier circuit in nanotechnology,'' \emph{Microelectronics
  Journal}, vol.~42, no.~8, pp. 973 -- 981, 2011. [Online]. Available:
  \url{http://www.sciencedirect.com/science/article/pii/S0026269211001194}
\BIBentrySTDinterwordspacing

\bibitem{Jamal}
L.~Jamal, M.~M. Rahman, and H.~M.~H. Babu, ``An optimal design of a fault
  tolerant reversible multiplier,'' in \emph{SOC Conference (SOCC), 2013 IEEE
  26th International}, Sept 2013, pp. 37--42.

\bibitem{Picca}
G.~{Florio} and D.~{Picca}, ``{Quantum implementation of elementary arithmetic
  operations},'' \emph{eprint arXiv:quant-ph/0403048}, Mar. 2004.

\bibitem{Kliuchnikov}
\BIBentryALTinterwordspacing
V.~Kliuchnikov, D.~Maslov, and M.~Mosca, ``Fast and efficient exact synthesis
  of single-qubit unitaries generated by clifford and t gates,'' \emph{Quantum
  Info. Comput.}, vol.~13, no. 7-8, pp. 607--630, Jul. 2013. [Online].
  Available: \url{http://dl.acm.org/citation.cfm?id=2535649.2535653}
\BIBentrySTDinterwordspacing

\bibitem{Kowada}
L.~A.~B. Kowada, R.~Portugal, and C.~M.~H. de~Figueiredo, ``Reversible
  karatsuba's algorithm,'' \emph{j-jucs}, vol.~12, no.~5, pp. 499--511, jun
  2006, \url
  $http://www.jucs.org/jucs\_12\_5/reversible\_karatsubas\_algorithm$.

\bibitem{Boss2}
\BIBentryALTinterwordspacing
H.~Thapliyal and N.~Ranganathan, ``Design of efficient reversible logic-based
  binary and bcd adder circuits,'' \emph{J. Emerg. Technol. Comput. Syst.},
  vol.~9, no.~3, pp. 17:1--17:31, Oct. 2013. [Online]. Available:
  \url{http://doi.acm.org/10.1145/2491682}
\BIBentrySTDinterwordspacing

\end{thebibliography}

%\begin{thebibliography}{1}

%\bibitem{IEEEhowto:kopka}
%H.~Kopka and P.~W. Daly, \emph{A Guide to {\LaTeX}}, 3rd~ed.\hskip 1em plus
 % 0.5em minus 0.4em\relax Harlow, England: Addison-Wesley, 1999.

%\end{thebibliography}

% biography section
% 
% If you have an EPS/PDF photo (graphicx package needed) extra braces are
% needed around the contents of the optional argument to biography to prevent
% the LaTeX parser from getting confused when it sees the complicated
% \includegraphics command within an optional argument. (You could create
% your own custom macro containing the \includegraphics command to make things
% simpler here.)
%\begin{biography}[{\includegraphics[width=1in,height=1.25in,clip,keepaspectratio]{mshell}}]{Michael Shell}
% where an .eps filename suffix will be assumed under latex, and a .pdf suffix
% will be assumed for pdflatex; or if you just want to reserve a space for
% a photo:

%\begin{biography}{Michael Shell}
%Biography text here.
%\end{biography}

%% if you will not have a photo at all:
%\begin{biographynophoto}{John Doe}
%Biography text here.
%\end{biographynophoto}

% insert where needed to balance the two columns on the last page
%\newpage

%\begin{biographynophoto}{Jane Doe}
%Biography text here.
%\end{biographynophoto}

% You can push biographies down or up by placing
% a \vfill before or after them. The appropriate
% use of \vfill depends on what kind of text is
% on the last page and whether or not the columns
% are being equalized.

%\vfill

% Can be used to pull up biographies so that the bottom of the last one
% is flush with the other column.
%\enlargethispage{-5in}

% that's all folks
\end{document}